\documentclass[12pt]{article}
\usepackage{hyperref}
\usepackage{amssymb,amsmath,amsthm}
\usepackage{enumerate}
\newcommand{\sss}{\setcounter{equation}{0}}
\renewcommand{\theequation}{\arabic{section}.\arabic{equation}}
\newtheorem{theorem}{THEOREM}[section]

\newtheorem{lemma}[theorem]{LEMMA}

\theoremstyle{definition}
\newtheorem{remark}[theorem]{REMARK}

\newcommand{\abs}[1]{\left|#1\right|}

\newcommand{\I}{{\rm i}}
\def\cprime{$'$}
\def\beq{\begin{equation}}
\def\ene{\end{equation}}

\DeclareMathOperator{\im}{Im}
\DeclareMathOperator{\dom}{Dom}

\DeclareMathOperator*{\res}{Res}

\begin{document}
\baselineskip=17 pt
\parskip 6 pt

\title{On the Two Spectra Inverse Problem for Semi-Infinite
  Jacobi Matrices  \thanks{ Research partially supported by
    Universidad Nacional Aut\'onoma de M\'exico under Project
    PAPIIT-DGAPA IN 105799, and by CONACYT under Project
    P42553­F.}}  \author{Luis O. Silva and Ricardo
  Weder\thanks{Fellow Sistema Nacional de Investigadores.}
  \\[4mm]
% Institution
Departamento de M\'{e}todos Matem\'{a}ticos y Num\'{e}ricos\\
Instituto de Investigaciones en Matem\'aticas Aplicadas y en Sistemas\\
Universidad Nacional Aut\'onoma de M\'exico\\
M\'exico, D.\,F.\, C.P. 04510
\\[4mm]
  \texttt{silva@leibniz.iimas.unam.mx}
  \\\texttt{weder@servidor.unam.mx}} \date{} \maketitle
\begin{center}
\begin{minipage}{5.75in}
  \centerline{{\bf Abstract}} \bigskip
  We present results on the unique reconstruction of a
  semi-infinite Jacobi operator from the spectra of the
  operator with two different boundary conditions.  This is
  the discrete analogue of the Borg-Marchenko theorem for
  Schr{\"o}dinger operators on the half-line. Furthermore, we
  give necessary and sufficient conditions for two real
  sequences to be the spectra of a Jacobi operator with
  different boundary conditions.
\end{minipage}
\end{center}
\newpage

%%%%%%%%%%%%%%%%
\section{Introduction}\sss
\label{sec:intro}
In the Hilbert space $l_2(\mathbb{N})$ let us single out the
dense subset $l_{fin}(\mathbb{N})$ of sequences which have a
finite number of non-zero elements.  Consider the operator $J$
defined for every $f=\{f_k\}_{k=1}^\infty$ in
$l_{fin}(\mathbb{N})$ by means of the recurrence relation
\begin{align}
  \label{eq:recurrence-coordinates}
  (Jf)_k &:= b_{k-1}f_{k-1} + q_k f_k + b_k
f_{k+1}\quad k \in \mathbb{N} \setminus \{1\}\\
  \label{eq:initial-coordinates}
 (Jf)_1 &:=  q_1 f_1 + b_1 f_2\,,
\end{align}
where, for every $n\in\mathbb{N}$, $b_n$ is positive, while
$q_n$ is real. $J$ is symmetric, therefore closable, and in
the sequel we shall consider the closure of $J$ and denote it
by the same letter.

Notice that we have defined the
Jacobi operator $J$ in such a way that
\begin{equation}
  \label{eq:jm-0}
  \begin{pmatrix}
    q_1 & b_1 & 0  &  0  &  \cdots
\\[1mm] b_1 & q_2 & b_2 & 0 & \cdots \\[1mm]  0  &  b_2  & q_3  &
b_3 &  \\
0 & 0 & b_3 & q_4 & \ddots\\ \vdots & \vdots &  & \ddots
& \ddots
  \end{pmatrix}
\end{equation}
is the matrix representation of $J$ with respect to the
canonical basis in $l_2(\mathbb{N})$ (we refer the reader to
\cite{MR1255973} for a discussion on matrix representation of
unbounded symmetric operators).

It is known that the symmetric operator $J$ has deficiency
indices $(1,1)$ or $(0,0)$ \cite[Chap. 4, Sec. 1.2]{MR0184042}
and \cite[Corollary 2.9]{MR1627806}. In the case $(1,1)$ we
can always define a linear set $D(g)\subset\dom (J^*)$
parametrized by $g\in\mathbb{R}\cup\{+\infty\}$ such that
\begin{equation*}
  J^*\upharpoonright D(g)=:J(g)
\end{equation*}
is a self-adjoint extension of $J$. Moreover, for any
self-adjoint extension (von Neumann extension) $\widetilde{J}$
of $J$, there exists a
$\widetilde{g}\in\mathbb{R}\cup\{+\infty\}$ such that
\begin{equation*}
  J(\widetilde{g})=\widetilde{J}\,,
\end{equation*}
\cite[Lemma 2.20]{MR1711536}. We shall
show later (see the Appendix) that $g$ defines a boundary
condition at infinity.

To simplify the notation, even in the case of deficiency
indices $(0,0)$, we shall use $J(g)$ to denote the operator
$J=J^*$. Thus, throughout the paper $J(g)$ stands either for a
self-adjoint extension of the nonself-adjoint operator $J$,
uniquely determined by $g$, or for the self-adjoint operator
$J$.

In what follows we shall consider the inverse spectral problem
for the self-adjoint operator $J(g)$.

It turns out that if $J\ne J^*$ (the case of indices $(1,1)$),
then for all $g\in\mathbb{R}\cup\{+\infty\}$ the Jacobi
operator $J(g)$ has discrete spectrum with eigenvalues of
multiplicity one, i.\,e., the spectrum consists of eigenvalues
of multiplicity one that can accumulate only at $\pm \infty$,
\cite[Lemma 2.19]{MR1711536}. Throughout this work we shall
always require that the spectrum of $J(g)$, denoted
$\sigma(J(g))$, be discrete, which is not an empty assumption
only for the case $J(g)=J$. Notice that the discreteness of
$\sigma(J(g))$ implies that $J(g)$ has to be unbounded.

For the Jacobi operators $J(g)$ one can define boundary
conditions at the origin in complete analogy to those of the
half-line Sturm-Liouville operator (see the Appendix).
Different boundary conditions at the origin define different
self-adjoint operators $J_h(g)$,
$h\in\mathbb{R}\cup\{+\infty\}$. $J_0(g)$ corresponds to the
Dirichlet boundary condition, while the operator $J_\infty(g)$
has Neumann boundary condition. If $J(g)$ has discrete
spectrum, the same is true for $J_h(g)$, $\forall
h\in\mathbb{R}\cup\{+\infty\}$ (for the case of $h$ finite see
Section 2 and for $h=\infty$, Section 4).

In this work we prove that a Jacobi operator $J(g)$ with
discrete spectrum is uniquely determined by
$\sigma(J_{h_1}(g))$, $\sigma(J_{h_2}(g))$, with
$h_1,h_2\in\mathbb{R}$ and $h_1 \ne h_2$, and either $h_1$ or
$h_2$. If $h_1$, respectively, $h_2$ is given, the
reconstruction method also gives $h_2$, respectively, $h_1$.
Saying that $J(g)$ is determined means that we can recover the
matrix (\ref{eq:jm-0}) and the boundary condition $g$ at
infinity, in the case of deficiency indices $(1,1)$.  We will
also establish (the precise statement is in Theorem
\ref{thm:suff-cond-two-gen}) that if two infinite real
sequences $\{\lambda_k\}_k$ and $\{\mu_k\}_k$ that can
accumulate only at $\pm \infty$ satisfy
  \begin{enumerate}[\ a)]
  \item  $\{\lambda_k\}_k$ and
  $\{\mu_k\}_k$ interlace, i.\,e., between two elements of a
  sequence there is one and only one element of the
  other. Thus, we assume below that $\lambda_k<\mu_k<\lambda_{k+1}$.
\item The series $\sum_k(\mu_k-\lambda_k)$ converges, so
  \begin{equation*}
   \sum_k(\mu_k-\lambda_k)=:\Delta < \infty\,.
\end{equation*}
By b) the product
$\displaystyle\prod_{k\ne n}
\frac{\mu_k-\lambda_n}{\lambda_k-\lambda_n}$ is convergent, so
define
  \begin{equation*}
    \tau_n^{-1}:=
   \frac{\mu_n-\lambda_n}{\Delta}
  \prod_{k\ne n}
\frac{\mu_k-\lambda_n}{\lambda_k-\lambda_n}\,.
  \end{equation*}
\item The sequence $\{\tau_n\}_n$ is such that, for
  $m=0,1,2,\dots$,
\begin{equation*}
   \sum_{k}\frac{\lambda_k^{2m}}{\tau_k}
    \quad\text{converges.}
\end{equation*}
\item For a sequence of complex numbers $\{\beta_k\}_k$,
  such that the series
  \begin{equation*}
    \sum_k\frac{\abs{\beta_k}^2}{\tau_k}
\quad\text{converges}
  \end{equation*}
and
\begin{equation*}
  \sum_{k}\frac{\beta_k\lambda_k^m}{\tau_k}=0\,,\qquad m=0,1,2,\dots 
\end{equation*}
it must hold true that $\beta_k=0$ for all $k$.
\end{enumerate}
Then, for any real number $h_1$, there exists a unique Jacobi
operator $J$, a unique $h_2>h_1$, and if $J \ne J^\ast$, a
unique $g\in\mathbb{R}\cup\{+\infty\}$, such that
$\sigma(J_{h_2}(g))=\{\lambda_k\}_k$ and
$\sigma(J_{h_1}(g))=\{\mu_k\}_k$. Moreover, we show that if
the sequences $\{\lambda_k\}_k$ and $\{\mu_k\}_k$ are the
spectra of a Jacobi operator $J(g)$ with two different
boundary conditions $h_1<h_2$ ($h_2\in\mathbb{R}$), then a),
b), c), d) hold for $\Delta=h_2-h_1$.

Necessary and sufficient conditions for two sequences to be
the spectra of a Jacobi operator $J(g)$ with Dirichlet and
Neumann boundary conditions are also given. Conditions b) and
c) differ in this case (see Section \ref{sec:d-n-cond}).

Our necessary and sufficient conditions give a
characterization of the spectral data for our two spectra
inverse problem.

Our proofs are constructive and they give a method for the
unique reconstruction of the operator $J$, the boundary
condition at infinity, $g$, and either $h_1$ or $h_2$.

The two-spectra inverse problem for Jacobi matrices has also
been studied in several papers
\cite{MR49:9676,MR499269,MR0221315,MR1643529}.  There are also
results on this problem in \cite{MR0190761}. We shall comment
on these results in the following sections.

The problem that we solve here is the discrete analogue of the
two-spectra inverse problem for Sturm-Liouville operators on
the half-line.  The classical result is the celebrated
Borg-Marchenko theorem \cite{3,4}. Let us briefly explain this
result.  Consider the self-adjoint Schr\"odinger operator,
\begin{equation}
  \label{eq:2-diff-expression}
  \mathcal{B} f=- f''(x) + Q(x)f(x)\,,\qquad
  x\in\mathbb{R}_+\,,
\end{equation}
where $Q(x)$ is real-valued and locally integrable on
$[0,\infty)$, and the following boundary condition at
zero is satisfied,
\begin{equation*}
   \cos\alpha f(0)+\sin\alpha f'(0)=0, \quad \alpha \in [0,\pi).
\end{equation*}
Moreover, the boundary condition at infinity, if any, is
considered fixed. Suppose that the spectrum is discrete for
one (and then for all) $\alpha$, and denote by
$\{\lambda_k(\alpha)\}_{k \in \mathbb N}$ the corresponding
eigenvalues.

The Borg-Marchenko theorem asserts that the sets
$\{\lambda_k(\alpha_1)\}_{k\in\mathbb{N}}$ and
$\{\lambda_k(\alpha_2)\}_{k\in\mathbb{N}}$ for some $\alpha_1
\ne \alpha_2$ uniquely determine $\alpha_1,\alpha_2$, and $Q$.
Thus, the differential expression and the boundary conditions
are determined by two spectra. Other results here are the
necessary and sufficient conditions for a pair of sequences to
be the eigenvalues of a  Sturm-Liouville equation with
different boundary conditions found by Levitan and Gasymov in
\cite{MR0162996}.

Other settings for two-spectra inverse problems can be found
in \cite{AW1,AW2,MR1329533}. A resonance inverse problem
for Jacobi matrices is considered in \cite{MR2164835}.  Recent
local Borg-Marchenko results for Schr\"odinger operators and
Jacobi matrices \cite{MR1754515,MR2091673} are also related to
the problem we discuss here.

Jacobi matrices appear in several fields of quantum mechanics
and condensed matter physics (see for example
\cite{MR883643}).

The paper is organized as follows. In Section 2 we present
some preliminary results that we need. In Section 3 we prove
our results of uniqueness, reconstruction, and necessary and
sufficient conditions (characterization) in the case where
$h_1$ and $h_2$ are real numbers. In Section 4 we obtain
similar results for the Dirichlet and Neumann boundary
conditions.  Finally, in the Appendix we briefly describe
--for the reader's convenience-- how the boundary conditions
are interpreted when $J$ is considered as a difference
operator.

\section{Preliminaries}\sss
\label{sec:preliminaries}
Let us denote by $\gamma$ the second order symmetric
difference expression (see (\ref{eq:recurrence-coordinates}),
(\ref{eq:initial-coordinates})) such that
$\gamma:f=\{f_k\}_{k\in \mathbb N}\mapsto \{(\gamma f)_k\}_{k \in
  \mathbb N}$, by
\begin{align}
  \label{eq:recurrence-spectral}
  (\gamma f)_k&:= b_{k-1}f_{k-1} + q_k f_k + b_kf_{k+1}\,,
  \quad k \in \mathbb{N} \setminus \{1\},\\
  \label{eq:initial-spectral}
 (\gamma f)_1&:= q_1 f_1 + b_1 f_2\,.
\end{align}
Then, it is proven in Section 1.1, Chapter 4 of
\cite{MR0184042} and in Theorem 2.7 of \cite{MR1627806} that
$$
\dom(J^\ast)=\{ f\in l_2(\mathbb N): \gamma f \in l_2(\mathbb
N)\},\quad J^\ast f= \gamma f,\qquad f \in \dom(J^\ast).  $$
The
solution of the difference equation,
\begin{equation}
  \label{2.1}
(\gamma f)= \zeta f\,,
\qquad \zeta \in \mathbb{C}\,,
\end{equation}
is uniquely determined if one gives $f_1=1$. For the elements
of this solution the following notation is standard
\cite[Chap. 1, Sec. 2.1]{MR0184042}
\begin{equation*}
  P_{n-1}(\zeta):=f_n\,,\qquad n\in\mathbb{N}\,,
\end{equation*}
where the polynomial $P_k(\zeta)$ (of degree $k$) is referred
to as the $k$-th orthogonal polynomial of the first kind
associated with the matrix (\ref{eq:jm-0}).

The sequence $\{P_k(\zeta)\}_{k=0}^\infty$ is not in
$l_{fin}(\mathbb{N})$ but it may happen that
\begin{equation}
 \label{eq:generalized-eigenvector}
  \sum_{k=0}^\infty\abs{P_k(\zeta)}^2<\infty\,,
\end{equation}
in which case $\zeta$ is an eigenvalue of $J^\ast$ and
$f(\zeta)$ the corresponding eigenvector. Since the eigenspace
is always one-dimensional, the eigenvalue of $J^\ast$ is of
multiplicity one . Moreover, since the (von Neumann) self-adjoint
extensions of $J, J(g),$ are restrictions of $J^\ast$, it
follows that the point spectrum of
$J(g),\,g\in\mathbb{R}\cup\{+\infty\}$, has multiplicity one.

The polynomials of the second kind
$\{Q_k(\zeta)\}_{k=0}^\infty$ associated with the matrix
(\ref{eq:jm-0}) are defined as the solutions of
\begin{equation*}
    b_{k-1}f_{k-1} + q_k f_k + b_kf_{k+1} = \zeta f_k\,,
\quad k \in \mathbb{N} \setminus \{1\}\,,
\end{equation*}
under the assumption that $f_1=0$ and $f_2=b_1^{-1}$. Then
\begin{equation*}
 Q_{n-1}(\zeta):=f_n\,,\qquad  n\in\mathbb{N}\,.
\end{equation*}
$Q_k(\zeta)$ is a polynomial of degree $k-1$.

By construction the Jacobi operator $J$ is a closed symmetric
operator. It is well known, \cite[Chap. 4, Sec.
1.2]{MR0184042} and \cite[Corollary 2.9]{MR1627806}, that this
operator has either deficiency indices $(1,1)$ or $(0,0)$. In
terms of the polynomials of the first kind, $J$ has deficiency
indices $(1,1)$ when
\begin{equation*}
  \sum_{k=0}^\infty\abs{P_k(\zeta)}^2<\infty\,,
  \quad\text{ for }\zeta\in\mathbb{C}\setminus\mathbb{R}
\end{equation*}
(this holds for all $\zeta\in\mathbb{C}\setminus\mathbb{R}$ if
and only if it holds for one
$\zeta\in\mathbb{C}\setminus\mathbb{R}$), and deficiency
indices $(0,0)$ otherwise. Since $J$ is closed, deficiency
indices $(0,0)$ mean that $J=J^*$. The symmetric operator $J$
with deficiency indices $(1,1)$ has always self-adjoint
extensions, which are restrictions of $J^*$. When studying the
self-adjoint extensions of $J$ in a more general context the
self-adjoint restrictions of $J^*$ are called von Neumann
self-adjoint extensions of $J$ \cite{MR1255973,MR1627806}. All
self-adjoint extensions considered in this paper are
von Neumann.

Let us now introduce a convenient way of parametrizing the
self-adjoint extensions of $J$ in the nonself-adjoint case.
We first define the Wronskian associated with $J$ for any pair
of sequences $\varphi=\{\varphi_k\}_{k=1}^\infty$ and
$\psi=\{\psi_k\}_{k=1}^\infty$ in $l_2(\mathbb{N})$ as follows
\begin{equation*}
  W_k(\varphi,\psi):=b_k(\varphi_k\psi_{k+1}-\psi_k\varphi_{k+1})\,,
  \quad k\in\mathbb{N}\,.
\end{equation*}
Now,  consider the sequences $v(g)=\{v_k(g)\}_{k=1}^\infty$
such that $\forall k\in\mathbb{N}$
\begin{equation}
  \label{eq:boundary-sequence}
  v_k(g):=P_{k-1}(0)+gQ_{k-1}(0)\,,
  \quad g\in\mathbb{R}
\end{equation}
and
\begin{equation}
  \label{eq:boundary-sequence-infty}
  v_k(+\infty):=Q_{k-1}(0)\,.
\end{equation}

All the self-adjoint extensions $J(g)$ of the nonself-adjoint
operator $J$ are restrictions of $J^*$ to the set \cite[Lemma
2.20]{MR1711536}
\begin{equation}
  \label{eq:beta-extensions-domain}
  \begin{split}
 D(g) &:= \bigl\{f=\{f_k\}_{k=1}^\infty\in\dom (J^*):\,
 \lim_{n\to\infty}W_n\bigl(v(g),f\bigr)=0\bigr\}=\\ &=
  \bigl\{f\in l_2(\mathbb{N}): \gamma f\in l_2(\mathbb{N}),
  \lim_{n\to\infty}W_n\bigl(v(g),f\bigr)=0\bigr\}\,.
\end{split}
\end{equation}
Different values of $g$ imply different self-adjoint
extensions. If $J$ is self-adjoint, we define
$J(g):=J$, for all $g\in\mathbb{R}\cup\{+\infty\}$; otherwise
$J(g)$ is a self-adjoint extension of $J$ uniquely determined
by $g$. We have defined the domains $D(g)$ in such a way that
$g$ defines a boundary condition at infinity (see the
Appendix).

It is worth mentioning that if $J\ne J^*$ then, for all
$g\in\mathbb{R}\cup\{+\infty\}$, $J(g)$ has discrete spectrum.
This follows from the fact that the resolvent of $J(g)$ turns
out to be a Hilbert-Schmidt operator
\cite[Lemma 2.19]{MR1711536}.

Let us now define the self-adjoint operator $J_h(g)$ by
\begin{equation*}
  J_h(g):=J(g)-h\langle\cdot,e_1\rangle e_1
  \,,\qquad  h \in\mathbb{R}\,,
\end{equation*}
where $\{e_k\}_{k=1}^\infty$ is the canonical basis in
$l_2(\mathbb{N})$ and $\langle\cdot,\cdot\rangle$ denotes the
inner product in this space. Clearly, $J_0(g)=J(g)$.

We define $J_\infty(g)$ as follows. First consider the space
$l_2\bigl((2,\infty)\bigr)$ of square summable sequences
$\{f_n\}_{n=2}^\infty$ and the sequence
$v(g)=\{v_k(g)\}_{k=1}^\infty$ given by
(\ref{eq:boundary-sequence}) and
(\ref{eq:boundary-sequence-infty}) with $g$ fixed. Let us
denote $J_\infty(g)$ the operator in
$l_2\bigl((2,\infty)\bigr)$ such that
\begin{equation*}
  J_\infty(g)f=\gamma f\,,
\end{equation*}
where $(\gamma f)_k$ is considered for any $k\ge 2$ and
$f_1=0$ in the definition of $(\gamma f)_2$, with domain given
by
\begin{equation*}
   \dom(J_\infty(g)):=\bigl\{f\in l_2\bigl((2,\infty)\bigr):
   \gamma f\in l_2\bigl((2,\infty)\bigr),
  \lim_{n\to\infty}W_n\bigl(v(g),f\bigr)=0\bigr\}\,.
\end{equation*}
Clearly, the matrix
\begin{equation*}
  \label{eq:jm-truncated}
  \begin{pmatrix}
    q_2 & b_2 & 0  &  0  &  \cdots
\\[1mm] b_2 & q_3 & b_3 & 0 & \cdots \\[1mm]  0  &  b_3  & q_4  &
b_4 &  \\
0 & 0 & b_4 & q_5 & \ddots\\ \vdots & \vdots &  & \ddots
& \ddots
  \end{pmatrix}\,,
\end{equation*}
which is our original matrix (\ref{eq:jm-0}) with the first
column and row removed, is the matrix representation of
$J_\infty(g)$ with respect to the canonical basis in
$l_2\bigl((2,\infty)\bigr)$.

It follows easily from the definition of $J_h(g)$ that if
$J(g)$ has discrete spectrum, the same is true for $J_h(g)$
($h\in\mathbb{R}\cup\{+\infty\}$). Indeed, for
$h\in\mathbb{R}$ this is a consequence of the invariance of
the essential spectrum --that is empty in our case-- under
a compact perturbation \cite{sch}. We shall show in Section
\ref{sec:d-n-cond} that it is also true that $J_\infty(g)$ has
discrete spectrum provided that $\sigma(J(g))$ is discrete.

For the self-adjoint operator $J_h(g)$, we can introduce the
right-continuous resolution of the identity $E_{J_h(g)}(t)$, such
that $J_h(g)=\int_\mathbb{R}tdE_{J_h(g)}(t)$. Let us define the
function $\rho(t)$ as follows:
\begin{equation}
\label{eq:spectral-measure}
  \rho(t):=\langle E_{J_h(g)}(t)e_1,e_1\rangle\,,
  \qquad t\in\mathbb{R}\,.
\end{equation}
Consider the function (see \cite{MR1643529} and \cite[Chap. 2,
Sec. 2.1]{MR1711536})
\begin{equation}
  \label{eq:m-weyl-as-resolvent}
  m_h(\zeta,g) :=
  \langle (J_h(g)-\zeta I)^{-1}e_1,e_1\rangle\,,
\qquad\zeta\not\in\sigma(J_h(g))\,.
\end{equation}
$m_h(\zeta,g)$ is called the Weyl $m$-function of $J_h(g)$. We
shall use below the simplified notation
$m(\zeta,g):=m_0(\zeta,g)$. The functions $\rho(t)$ and
$m_h(\zeta,g)$ are related by the Stieltjes transform (also
called Borel transform):
\begin{equation*}
  m_h(\zeta,g) =\int_{\mathbb{R}}\frac{d\rho(t)}{t-\zeta}\,.
\end{equation*}
It follows from the definition  that the Weyl $m$-function
is a Herglotz function, i.\,e.,
\begin{equation*}
  \frac{\im m_h(\zeta,g) }{\im \zeta}>0\,,\qquad\im\zeta>0\,.
\end{equation*}

Using the Neumann expansion for the resolvent
(cf.\cite[Chap. 6, Sec. 6.1]{MR1711536})
\begin{equation*}
  (J_h(g)-\zeta I)^{-1}=
  -\sum_{k=0}^{N-1}\frac{(J_h(g))^k}{\zeta^{k+1}}
  +\frac{(J_h(g))^N}{\zeta^{N}}
  (J_h(g)-\zeta I)^{-1}\,,
\end{equation*}
where $\zeta\in\mathbb{C}\setminus\sigma(J(g))$,
one can easily obtain the following asymptotic formula
\begin{equation}
  \label{eq:m-asympt}
  m_h(\zeta,g)=-\frac{1}{\zeta}-\frac{q_1-h}{\zeta^2}
  -\frac{b_1^2+(q_1-h)^2}{\zeta^3}
  +O(\zeta^{-4})\,,
\end{equation}
as $\zeta\to\infty$ ($\im \zeta\ge \epsilon$, $\epsilon>0$).

An important result in the theory of Jacobi operators is the
fact that $m(\zeta,g)$ completely determines $J(g)$ (the same
is of course true for the pair $m_h(\zeta,g)$ and $J_h(g)$).
There are two ways for recovering the operator from the Weyl
$m$-function.  One way consists in obtaining first $\rho(t)$ from
$m(\zeta,g)$ by means of the inverse
Stieltjes transform (cf. \cite[Appendix B]{MR1711536}), namely,
\begin{equation*}
  \rho(b)-\rho(a)=\lim_{\delta\downarrow 0}\lim_{\epsilon\downarrow 0}
  \frac{1}{\pi}\int_{a+\delta}^{b+\delta}\, \left(\im m(x+\I\epsilon,g)\right)
  dx\,.
\end{equation*}
The function $\rho$ is such that all the moments of the
corresponding measure are finite \cite{MR0184042,MR1627806}.
Hence, all the elements of the sequence $\{t^k\}_{k=0}^\infty$
are in $L_2(\mathbb{R},d\rho)$ and one can apply, in this
Hilbert space, the Gram-Schmidt procedure of
orthonormalization to the sequence $\{t^k\}_{k=0}^\infty$.
One, thus, obtains a sequence of polynomials
$\{P_k(t)\}_{k=0}^\infty$ normalized and orthogonal in
$L_2(\mathbb{R},d\rho)$. These polynomials satisfy a three term
recurrence equation \cite{MR1627806}
\begin{align}
\label{eq:favard-system1}
      tP_{k-1}(t) &= b_{k-1}P_{k-2}(t) +  q_k
      P_{k-1}(t) +  b_k P_k(t)
\quad k \in \mathbb{N} \setminus \{1\}\\
\label{eq:favard-system2}
 tP_0(t) &=  q_1 P_0(t) +  b_1 P_1(t)\,,
\end{align}
where all the coefficients $b_k$ ($k\in\mathbb{N}$) turn out
to be positive and $q_k$ ($k\in\mathbb{N}$) are real numbers.
The system (\ref{eq:favard-system1}) and
(\ref{eq:favard-system2}) defines a matrix which is the matrix
representation of $J$. We shall refer to this procedure for
recovering $J$ as the method of orthogonal polynomials. The
other method for determining $J$ from $m(\zeta,g)$ was
developed in \cite{MR1616422} (see also \cite{MR1643529}). It
is based on the asymptotic behavior of $m(\zeta,g)$ and the
Ricatti equation \cite{MR1643529},
\begin{equation}
  \label{eq:ricatti}
    b_n^2 m^{(n)}(\zeta,g)=
    q_n-\zeta-\frac{1}{m^{(n-1)}(\zeta,g)}\,,\quad n\in\mathbb{N}\,,
\end{equation}
where $m^{(n)}(\zeta,g)$ is the Weyl $m$-function of the
Jacobi operator associated with the matrix (\ref{eq:jm-0})
with the first $n$ columns and $n$ rows removed.

After obtaining the matrix representation of $J$, one can
easily obtain the boundary condition at infinity which defines
the domain of $J(g)$ in the nonself-adjoint case. Indeed, take
an eigenvalue, $\lambda$, of $J(g)$, i.\,e., $\lambda$ is a
pole of $m(\zeta,g)$. Since the corresponding eigenvector
$f(\lambda)=\{f_k(\lambda)\}_{k=1}^\infty$ is in $\dom
(J(g))$, it must be that
\begin{equation*}
  \lim_{n\to\infty}W_n\bigl(v(g),f(\lambda)\bigr)=0\,.
\end{equation*}
This implies that either
$\lim_{n\to\infty}
W_n\bigl(\{Q_{k-1}(0)\}_{k=1}^\infty,f(\lambda)\bigr)=0$,
which means that $g=+\infty$, or
\begin{equation*}
  g=-\frac
  {\lim_{n\to\infty}W_n\bigl(\{P_{k-1}(0)\}_{k=1}^\infty,f(\lambda)\bigr)}
  {\lim_{n\to\infty}W_n\bigl(\{Q_{k-1}(0)\}_{k=1}^\infty,f(\lambda)\bigr)}
  \,.
\end{equation*}

If the spectrum of $J_h(g)$ is discrete, say
$\sigma(J_h(g))=\{\lambda_k\}_k$, the function $\rho(t)$ defined
by (\ref{eq:spectral-measure}) can be written as follows
\begin{equation*}
   \rho(t)=\sum_{\lambda_k\le t}\frac{1}{\alpha_k}\,,
\end{equation*}
where the coefficients $\{\alpha_k\}_k$ are called the
normalizing constants and are given by
\begin{equation}
  \label{eq:def-normalizing}
  \alpha_n=\sum_{k=0}^\infty\abs{P_k(\lambda_n)}^2\,.
\end{equation}
Thus, $\sqrt{\alpha_n}$ equals the $l_2$ norm of the
eigenvector $f(\lambda_n):=\{P_k(\lambda_n)\}_{k=0}^\infty$
corresponding to $\lambda_n$. The eigenvector $f(\lambda_n)$
is normalized in such a way that $f_1(\lambda_n)=1$.

Clearly,
\begin{equation}
  \label{eq:sum-normalizing-constants}
  1=\langle e_1,e_1\rangle=\int_\mathbb{R}d\rho=
  \sum_{k}\frac{1}{\alpha_k}\,.
\end{equation}
The Weyl $m$-function in this case is given by
\begin{equation}
  \label{eq:m-normalizing}
  m_h(\zeta,g)=\sum_{k}\frac{1}{\alpha_k(\lambda_k-\zeta)}\,.
\end{equation}
From this we have that
\begin{equation*}
  (\lambda_n-\zeta)m_h(\zeta,g) =
  (\lambda_n-\zeta)\sum_k\frac{1}{\alpha_k(\lambda_k-\zeta)}=
  \sum_{k\ne
  n}\frac{\lambda_n-\zeta}{\alpha_k(\lambda_k-\zeta)} +
  \frac{1}{\alpha_n}\,.
\end{equation*}
Therefore,
\begin{equation}
  \label{eq:normalizing-const-formula}
  \alpha_n^{-1}=\lim_{\zeta\to\lambda_n}(\lambda_n-\zeta)m(\zeta,g)
  =-\res_{\zeta=\lambda_n}m(\zeta,g)\,.
\end{equation}

Let us now introduce an appropriate way for enumerating
sequences that we shall use. Consider a pair of infinite real
sequences $\{\lambda_k\}_k$ and $\{\mu_k\}_k$ that have no
finite accumulation points and that interlace, i.\,e., between
two elements of one sequence there is one and only one element
of the other. We use $M$, a subset of $\mathbb{Z}$ to be
defined below, for enumerating the sequences as follows
\begin{equation}
  \label{eq:enum-zeros-poles}
  \forall k\in M\quad \lambda_k<\mu_k<\lambda_{k+1},
\end{equation}
where
\begin{enumerate}[\ a)]
\item If $\inf_k\{\lambda_k\}_k=-\infty$ and
  $\sup_k\{\lambda_k\}_k=\infty$,
  \begin{equation}
    \label{eq:unbounded-case}
    M:=\mathbb{Z}\quad\text{and we require}\quad
\mu_{-1} < 0 < \lambda_1\,.
  \end{equation}
\item
If $0<\sup_k\{\lambda_k\}_k<\infty$,
\begin{equation}
  \label{eq:bounded-above-positive}
  M:= \{k\}_{k=-\infty}^{k_{\max}}\,,
  \ (k_{\max}\ge 1)\quad\text{and we require}\quad
\mu_{-1} < 0 < \lambda_1\,.
\end{equation}
\item
If $\sup_k\{\lambda_k\}_k \leq 0$,
\begin{equation}
  \label{eq:bounded-above-0}
  M:=\{k\}_{k=-\infty}^0\,.
\end{equation}
\item
If $\inf_k\{\mu_k\}_k\geq 0$,
\begin{equation}
  \label{eq:bounded-below-0}
  M:=\{k\}_{k=0}^\infty\,.
\end{equation}
\item If $-\infty<\inf_k\{\mu_k\}_k < 0$,
  \begin{equation}
    \label{eq:bounded-below-neg}
    M:=\{k\}_{k=k_{\min}}^{\infty}\,,
  \ (k_{\min}\le -1)\quad\text{and we require}\quad
\mu_{-1} < 0 < \lambda_1\,.
  \end{equation}
\end{enumerate}
Notice that, by this convention for enumeration, the
only elements of $\{\lambda_k\}_{k\in M}$ and $\{\mu_k\}_{k\in
  M}$ allowed to be zero are $\lambda_0$ or $\mu_0$.
\section{Rank one
  perturbations with finite coupling constants}\sss
\label{sec:sep-conditions}
In this section we consider a pair of operators $J_{h_1}(g)$
and $J_{h_2}(g)$, where $h_1,h_2\in \mathbb{R}$, that is, rank
one perturbations of the Jacobi operator $J(g)$ with finite
coupling constants.

\subsection{Recovering the matrix from two spectra}
\label{sec:recov-matr-from-1}

Let $g\in\mathbb{R}\cup\{+\infty\}$ be fixed. Since $J_h(g)$
is a rank one perturbation of $J(g)$, the domain of $J(g)$
coincides with the domain of $J_h(g)$ for all
$h\in\mathbb{R}$. Moreover, since the perturbation is analytic in
$h$, the multiplicity-one eigenvalues, $\lambda_k(h)$, and the
corresponding eigenvectors, are
analytic functions of $h$ \cite{ka}.
\begin{lemma}
  \label{lem:halilova1}
  Let $\{\lambda_k(h)\}_k$ be the set of eigenvalues of $J_h(g)$
  ($h\in\mathbb{R}$). For a fixed $k$ the
  following holds
\begin{equation}
  \label{eq:lem-halilova}
  \frac{d}{dh}\lambda_k(h)=-\frac{1}{\alpha_k(h)}\,,
\end{equation}
where $\alpha_k(h)$ is the normalizing constant corresponding
to $\lambda_k(h)$.
\end{lemma}
\begin{proof}
  For the sake of simplifying the formulae, we write $J_h$ and
  $\lambda(h)$ instead of $J_h(g)$ and $\lambda_k(h)$,
  respectively ($k$ is fixed). Let us denote by $f(h)$ the
  eigenvector of $J_h$ corresponding to $\lambda(h)$.  Take
  any $\delta>0$, taking into account that $\dom
  (J_{h+\delta})=\dom (J_h)$ and that $J_h$ is symmetric for
  any $h\in\mathbb{R}$, we have that
  \begin{gather*}
  (\lambda(h+\delta)-\lambda(h))
  \langle f(h+\delta),f(h)\rangle=\\
  \langle J_{h+\delta}f(h+\delta),f(h)\rangle -
  \langle f(h+\delta),J_{h}f(h)\rangle=\\=
  \langle (J_{h+\delta}-J_{h}+J_{h})f(h+\delta),f(h)\rangle -
  \langle f(h+\delta),J_{h}f(h)\rangle =\\=
 \langle (J_{h+\delta}-J_{h})f(h+\delta),f(h)\rangle = -\delta\,.
  \end{gather*}
Therefore,
\begin{equation*}
  \lim_{\delta\to
  0}\frac{\lambda(h+\delta)-\lambda(h)}{\delta}=
 -\lim_{\delta\to
  0}\frac{1}{\langle f(h+\delta),f(h)\rangle}=
  -\frac{1}{\alpha_k(h)}\,.
\end{equation*}\\
\end{proof}
The cornerstone of our analysis below is the Weyl
$m$-function. Let us establish the relation between
$m_h(\zeta,g)$ and $m(\zeta,g)$. Consider the second
resolvent identity \cite{MR566954}:
\begin{equation}
  \label{eq:second-resolvent-identity}
  (J_h(g)-\zeta I)^{-1}-(J(g)-\zeta I)^{-1}=
  (J(g)-\zeta I)^{-1}(J(g)-J_h(g))(J_h(g)-\zeta I)^{-1}\,,
\end{equation}
where $\zeta\in\mathbb{C}\setminus\{\sigma(J(g))\cup\sigma(J_h(g))\}$.
Then, for $h\in\mathbb{R}$,
\begin{equation*}
  \begin{split}
    m_h(\zeta,g)-m(\zeta,g)&=
    \langle\left((J_h(g)-\zeta I)^{-1}-(J(g)-\zeta
    I)^{-1}\right)e_1,e_1\rangle \\ &=
  \big\langle(J(g)-\zeta I)^{-1}
    (h\langle\cdot,e_1\rangle e_1)
    (J_h(g)-\zeta I)^{-1}e_1,e_1\big\rangle \\ &=
     \big\langle
    h\langle(J_h(g)-\zeta I)^{-1}e_1,e_1\rangle
    (J(g)-\zeta I)^{-1}
    e_1, e_1\big\rangle \\ &=
    h m_h(\zeta,g)m(\zeta,g)\,.
  \end{split}
\end{equation*}
Hence,
\begin{equation}
  \label{eq:m-two-boundaries}
  m_h(\zeta,g)=\frac{m(\zeta,g)}{1-hm(\zeta,g)}\,.
\end{equation}
\begin{remark}
  \label{rem:zeros-poles-m-h}
  If $J(g)$ has discrete spectrum, then $m(\zeta,g)$ is
  meromorphic and, by (\ref{eq:m-two-boundaries}), so is
  $m_h(\zeta,g)$. The poles of $m_h(\zeta,g)$ are the
  eigenvalues of $J_h(g)$. Since the poles of the denominator
  and numerator in (\ref{eq:m-two-boundaries}) coincide,
  assuming that $h\ne 0$, the
  poles of $m_h(\zeta,g)$ are given by the zeros of
  $1-hm(\zeta,g)$ and the zeros of $m_h(\zeta,g)$ by the zeros
  of $m(\zeta,g)$. Thus, $J_{h_1}(g)$ and $J_{h_2}(g)$
  have different eigenvalues, provided that $h_1\ne h_2$.
\end{remark}
\begin{theorem}
\label{uniq.1}
  \label{th:recon-by-levin-gen}
  Consider the Jacobi operator $J(g)$ with discrete spectrum.
  The sequences $\{\mu_k\}_k=\sigma(J_{h_1}(g))$ and
  $\{\lambda_k\}_k=\sigma(J_{h_2}(g)), h_1 \ne h_2$, together
  with $h_1$ (respectively, $h_2$) uniquely determine the
  operator $J$, $h_2,$ (respectively, $h_1$) and, if $J\ne
  J^*$, the boundary condition $g$ at infinity.
\end{theorem}
\begin{proof}
  Without loss of generality we can assume that $h_1 < h_2$.
  Consider the Weyl $m$-function $m(\zeta,g)$ of the operator
  $J(g)$.  Let us define the function
  \begin{equation}
    \label{eq:teschl-function}
    \mathfrak{m}(\zeta,g)=\frac{m_{h_2}(\zeta,g)}{m_{h_1}(\zeta,g)}\,,
    \qquad\zeta\in\mathbb{C}\setminus\mathbb{R}\,.
  \end{equation}
  Notice first that the zeros of $\mathfrak{m}(\zeta,g)$ are
  the eigenvalues of $J_{h_1}(g)$ while the poles of
  $\mathfrak{m}(\zeta,g)$ are the eigenvalues of $J_{h_2}(g)$.
  This follows from Remark~\ref{rem:zeros-poles-m-h} and
  (\ref{eq:teschl-function}).  Let us now show that
  $\mathfrak{m}(\zeta,g)$ is a Herglotz or an
  anti-Herglotz function. Indeed, since $m(\zeta,g)$ is
  Herglotz, then
  \begin{equation}
    \label{eq:m-h-herglotz}
     \mathfrak{m}(\zeta,g)=\frac{1-h_1m}{1-h_2m}=
     1+\frac{-1}{\frac{h_2}{h_2-h_1}+
       \frac{-1}{(h_2-h_1)m(\zeta,g)}}\,.
 \end{equation}
 Therefore, $\mathfrak{m}(\zeta,g)$ is Herglotz or
 anti-Herglotz depending on the sign of $h_2-h_1$.  Recall
 that if a function $f$ is Herglotz, then, $-\frac{1}{f}$ is
 also Herglotz. Since $h_2-h_1>0$, $\mathfrak{m}(\zeta,g)$ is
 a Herglotz function.

 Thus, the zeros $\{\mu_k\}_k$ of $ \mathfrak{m}(\zeta,g)$ and
 its poles $\{\lambda_k\}_k$ interlace. Let us use the
 convention
 (\ref{eq:enum-zeros-poles})--(\ref{eq:bounded-below-neg}) for
 enumerating the zeros and poles of $\mathfrak{m}(\zeta,g)$.
 By this convention, if the sequence $\{\lambda_k\}_k$ (or
 $\{\mu_k\}_k$) is bounded from below, the least of all zeros
 is greater than the least of all poles, while, if
 $\{\lambda_k\}_k$ is bounded from above, the greatest of all
 poles is less than the greatest of all zeros. It is easy to
 verify, using for instance (\ref{eq:lem-halilova}), that this
 is what we have for the zeros and poles of
 $\mathfrak{m}(\zeta,g)$ when $J(g)$ is semi-bounded.

 According to \cite[Chap. 7, Sec.1, Theorem 1]{MR589888}, the
 meromorphic Herglotz function $\mathfrak{m}(\zeta,g)$, with
 its zeros and poles enumerated as convened, can be written as
 follows
\begin{equation}
  \label{eq:levin-herglotz-gen}
   \mathfrak{m}(\zeta,g)= C \frac{\zeta-\mu_0}{\zeta-\lambda_0}
  \sideset{}{'}\prod_{k\in M} \left(1-\frac{\zeta}{\mu_k}\right)
  \left(1-\frac{\zeta}{\lambda_k}\right)^{-1}\,,\qquad C>0\,,
\end{equation}
where the prime in the infinite product means that it does not
include the factor $k=0$.

From the asymptotic behavior of $m(\zeta,g)$, given by
(\ref{eq:m-asympt}), one easily obtains that, as
$\zeta\to\infty$ with $\im \zeta\ge\epsilon$ ($\epsilon>0$),
\begin{equation}
  \label{eq:M-asympt}
  \mathfrak{m}(\zeta,g)=1+(h_1-h_2)\zeta^{-1}
  +(h_1-h_2)(q_1 -h_2)\zeta^{-2}+O(\zeta^{-3})\,.
\end{equation}
Therefore,
\begin{equation*}
   \lim_{\substack{\zeta\to\infty \\ \im \zeta\ge\epsilon}}
    \mathfrak{m}(\zeta,g)=1\,.
\end{equation*}
Then, using (\ref{eq:levin-herglotz-gen}), we have
\begin{equation}
  \label{eq:c-deter-2-spect-gen}
  C^{-1}=\lim_{\substack{\zeta\to\infty \\
      \im \zeta\ge\epsilon}}
\sideset{}{'}\prod_{k\in M} \left(1-\frac{\zeta}{\mu_k}\right)
  \left(1-\frac{\zeta}{\lambda_k}\right)^{-1}
  \,,\qquad \epsilon>0\,.
\end{equation}
Thus, $\mathfrak{m}(\zeta,g)$ is completely determined by the
spectra $\sigma(J_{h_1}(g))$ and $\sigma(J_{h_2}(g))$. Having
found $\mathfrak{m}(\zeta,g)$, we can determine $h_2,$
respectively, $h_1$, by means of (\ref{eq:M-asympt}). Hence,
from (\ref{eq:m-h-herglotz}) one obtains $m(\zeta,g)$ and,
using the methods introduced in the preliminaries, $J$ is
uniquely determined. In the case when $J\ne J^*$, we can also
find the boundary condition $g$ at infinity as indicated in
Section 2.\\
\end{proof}

In \cite{MR1643529} (see also \cite{MR0190761}) it is proven
that the discrete spectra of $J_{h_1}(g)$ and $J_{h_2}(g)$,
together with $h_1$ and $h_2$ uniquely determine $J$ and the
boundary condition $g$ in the $(1,1)$ case. Our result shows
that it is not necessary to know both $h_1$ and $h_2$, one of
them is enough.

It turns out that if one knows the spectra
$\sigma(J_{h_1}(g))$ and $\sigma(J_{h_2}(g))$ together with
$q_1$, the first element of the matrix's main diagonal, it is
possible to recover uniquely the matrix, the boundary
conditions $h_1$, $h_2$ and the boundary condition at
infinity, $g$, if any. Indeed, the term of order $\zeta^{-1}$ in
the asymptotic expansion of $\mathfrak{m}(\zeta,g)$
(\ref{eq:M-asympt}) determines $h_1-h_2$. Since the coefficient
of $\zeta^{-2}$ term is $(h_1-h_2)(q_1-h_2)$, if we know $q_1$
one finds $h_2$, and then $h_1$.

\subsection{Necessary and Sufficient conditions }
\label{sec:suff-cond-two-gen}
\begin{theorem}
\label{thm:suff-cond-two-gen}
Given $h_1\in\mathbb{R}$ and two infinite sequences of real
numbers $\{\lambda_k\}_k$ and $\{\mu_k\}_k$ without finite
points of accumulation, there is a unique real $ h_2 > h_1$, a
unique operator $J(g)$, and if $J \ne J^\ast$ also a unique $g
\in \mathbb R \cup \{+\infty\}$, such that, $\{\mu_k\}_k
=\sigma(J_{h_1}(g))$ and $\{\lambda_k\}_k =\sigma(J_{h_2}(g))$
if and only if the following conditions are satisfied.
\begin{enumerate}[\ a)]
\item $\{\lambda_k\}_k$ and $\{\mu_k\}_k$ interlace and, if
  $\{\lambda_k\}_k$ is bounded from below,
  $\min_k\{\mu_k\}_k>\min_k\{\lambda_k\}_k$, while if
  $\{\lambda_k\}_k$ is bounded from above,
  $\max_k\{\lambda_k\}_k<\max_k\{\mu_k\}_k$. So we use below
  the convention
  (\ref{eq:enum-zeros-poles})--(\ref{eq:bounded-below-neg}) for
  enumerating the sequences.
    \label{interlace-sufficient}
\item The following series converges
  \begin{equation*}
   \sum_{k\in M}(\mu_k-\lambda_k)=\Delta<\infty\,.
\end{equation*}\label{sum-spectr-sufficient}
By condition \ref{sum-spectr-sufficient}) the product
$\displaystyle\prod_{\substack{k\in M\\k\ne n}}
\frac{\mu_k-\lambda_n}{\lambda_k-\lambda_n}$ is convergent, so
define
  \begin{equation}
    \label{eq:tau-def-1}
    \tau_n^{-1}:=
   \frac{\mu_n-\lambda_n}{\Delta}
  \prod_{\substack{k\in M\\k\ne n}}
\frac{\mu_k-\lambda_n}{\lambda_k-\lambda_n}\,,
  \quad \forall n\in M\,.
  \end{equation}
\item The sequence $\{\tau_n\}_{n\in M}$
is such that, for $m=0,1,2,\dots$, the series
\begin{equation*}
    \sum_{k\in M}\frac{\lambda_k^{2m}}{\tau_k}
    \quad\text{converges.}
\end{equation*}
\label{finite-moments-sufficient}
\item If a sequence of complex numbers $\{\beta_k\}_{k\in M}$
  is such that the series
  \begin{equation*}
    \sum_{k\in
      M}\frac{\abs{\beta_k}^2}{\tau_k}
\quad\text{converges}
  \end{equation*}
and, for $m=0,1,2,\dots$,
\begin{equation*}
  \sum_{k\in
    M}\frac{\beta_k\lambda_k^m}{\tau_k}=0\,,
\end{equation*}
then $\beta_k=0$ for all $k\in M$.
\label{density-poly-sufficient}
  \end{enumerate}
\end{theorem}
\begin{proof}
  We first prove that if $\{\lambda_k\}_k$ and $\{\mu_k\}_k$
  are the spectra of $J_{h_2}(g)$ and $J_{h_1}(g)$, with $ h_2
  > h_1$, then \emph{\ref{interlace-sufficient}}),
  \emph{\ref{sum-spectr-sufficient}}),
  \emph{\ref{finite-moments-sufficient}}), and
  \emph{\ref{density-poly-sufficient}}) hold true. The
  condition \emph{\ref{interlace-sufficient}}) follows
  directly from the proof of the previous theorem. To prove
  that \emph{\ref{sum-spectr-sufficient}}) holds, observe that
  (\ref{eq:lem-halilova}) implies
  \begin{equation*}
    \mu_k-\lambda_k=
    \int_{h_1}^{h_2}\frac{dh}{\alpha_k(h)}\,.
  \end{equation*}
  Consider a sequence $\{M_n\}_{n=1}^\infty$ of subsets of
  $M$, such that $M_n\subset M_{n+1}$ and $\cup_nM_n=M$, then,
  using (\ref{eq:sum-normalizing-constants}), we have
  \begin{equation*}
  s_n:=\sum_{k\in M_n}(\mu_k-\lambda_k)=
  \sum_{k\in M_n}\int_{h_1}^{h_2}\frac{dh}{\alpha_k(h)}=
  \int_{h_1}^{h_2}\sum_{k\in M_n}\frac{dh}{\alpha_k(h)}\le
  h_2-h_1\,.
  \end{equation*}
  The sequence $\{s_n\}_{n=1}^\infty$ is then convergent and
  clearly
  \begin{equation*}
  \sum_{k\in M}(\mu_k-\lambda_k)=\lim_{n\to\infty}s_n=h_2-h_1\,.
  \end{equation*}
  Thus, $\Delta=h_2-h_1$.

  The convergence of the series in
  \emph{\ref{sum-spectr-sufficient}}) allows us to write
  (\ref{eq:c-deter-2-spect-gen}) as follows
  \begin{equation*}
    C^{-1}=
    \sideset{}{'}\prod_{k\in M}\frac{\lambda_k}{\mu_k}
    \lim_{\substack{\zeta\to\infty \\
        \im \zeta\ge\epsilon}}
    \sideset{}{'}\prod_{k\in M}
    \frac{\mu_k-\zeta}{\lambda_k-\zeta}
    \,,\qquad \epsilon>0\,.
  \end{equation*}
  Now, using again \emph{\ref{sum-spectr-sufficient}}),
  it easily follows that for any $\epsilon>0$
  \begin{equation*}
    \lim_{\substack{\zeta\to\infty \\
        \im \zeta\ge\epsilon}}
    \sideset{}{'}\prod_{k\in M}
    \frac{\mu_k-\zeta}{\lambda_k-\zeta}=
    \lim_{\substack{\zeta\to\infty \\
        \im \zeta\ge\epsilon}}
    \sideset{}{'}\prod_{k\in M}
    \left(1+\frac{\mu_k-\lambda_k}{\lambda_k-\zeta}\right)=1\,.
  \end{equation*}
  Thus, $C=\sideset{}{'}\prod_{k\in M}\mu_k/\lambda_k$ and by
  (\ref{eq:levin-herglotz-gen}),
  \begin{equation}
   \label{eq:short}
   \mathfrak{m}(\zeta,g)=\prod_{k\in M}
   \frac{\mu_k-\zeta}{\lambda_k-\zeta}\,.
  \end{equation}
  Let us now find formulae for the normalizing constants in
  terms of the sets of eigenvalues for different boundary
  conditions. By (\ref{eq:normalizing-const-formula}),
  \begin{equation*}
    \alpha_n^{-1}(h_2,g)=\lim_{\zeta\to\lambda_n}
    (\lambda_n-\zeta)m_{h_2}(\zeta,g)\,.
  \end{equation*}
  Using the second resolvent identity, as we did to obtain
  (\ref{eq:m-two-boundaries}), we
  have that
  \begin{equation*}
    m_{h_1}(\zeta,g)=
    \frac{m_{h_2}(\zeta,g)}{1-(h_1-h_2)m_{h_2}(\zeta,g)}\,.
  \end{equation*}
  Therefore,
  \begin{equation}
    \label{eq:m-gen-m-h}
    \mathfrak{m}(\zeta,g)=
    \frac{m_{h_2}(\zeta,g)}{m_{h_1}(\zeta,g)}
    = 1-(h_1-h_2)m_{h_2}\,,
    \qquad\zeta\in\mathbb{C}\setminus\mathbb{R}\,.
  \end{equation}
  Then, the normalizing constants are given by
  \begin{equation*}
    \alpha_n^{-1}(h_2,g)=\lim_{\zeta\to\lambda_n}
    (\lambda_n-\zeta)\frac{\mathfrak{m}(\zeta,g) -1}{h_2-h_1}=
    \frac{1}{h_2-h_1}\lim_{\zeta\to\lambda_n}
    (\lambda_n-\zeta)\mathfrak{m}(\zeta,g)\,.
  \end{equation*}
  Now,
  \begin{equation}
    \label{eq:res-m-computation}
  \begin{split}
    \lim_{\zeta\to\lambda_n}(\lambda_n-\zeta)\mathfrak{m}(\zeta,g)&=
    \lim_{\zeta\to\lambda_n}(\lambda_n-\zeta) \prod_{k\in M}
    \frac{\mu_k-\zeta}{\lambda_k-\zeta}=\\
    &=(\mu_n-\lambda_n) \prod_{\substack{k\in M\\k\ne n}}
    \frac{\mu_k-\lambda_n}{\lambda_k-\lambda_n}\,.
  \end{split}
  \end{equation}
  Hence,
  \begin{equation}
    \label{eq:n-normalizing-spect-gen}
    \alpha_n^{-1}(h_2,g)=
    \frac{\mu_n-\lambda_n}{h_2-h_1}
    \prod_{\substack{k\in M\\k\ne n}}
    \frac{\mu_k-\lambda_n}{\lambda_k-\lambda_n}\,.
  \end{equation}
  Notice that, since $\Delta=h_2-h_1$, it follows from
  (\ref{eq:n-normalizing-spect-gen}) that $\tau_n=\alpha_n$
  for all $n\in M$. Hence the spectral function $\rho$ of the
  self-adjoint extension $J_{h_2}(g)$ is given by the
  expression $\rho(t)=\sum_{\lambda_k\le t}\tau_k^{-1}$. Thus
  \emph{\ref{finite-moments-sufficient}}) follows from the
  fact that all the moments of $\rho$ are finite
  \cite{MR0184042,MR1627806}. Similarly,
  \emph{\ref{density-poly-sufficient}}) stems from the density
  of polynomials in $L_2(\mathbb{R},d\rho)$, which takes place
  since $\rho$ is $N$-extremal \cite{MR0184042},
  \cite[Proposition 4.15]{MR1627806}.

  We now prove that conditions
  \emph{\ref{interlace-sufficient}}),
  \emph{\ref{sum-spectr-sufficient}}),
  \emph{\ref{finite-moments-sufficient}}), and
  \emph{\ref{density-poly-sufficient}}) are sufficient.
  Let $\{\lambda_k\}_k$ and $\{\mu_k\}_k$ be sequences as in
\emph{\ref{interlace-sufficient}}) and
  \emph{\ref{sum-spectr-sufficient}}). Then,
\begin{equation}
  \label{eq:prod-posit-gen}
0<\prod_{\substack{k\in M\\k\ne n}}
\frac{\mu_k-\lambda_n}{\lambda_k-\lambda_n}<\infty\,.
\end{equation}
The convergence of this product allows us to define the sequence
of numbers $\{\tau_n\}_{n\in M}$. Observe that for all $n\in
M$, $\tau_n>0$. Indeed, $\Delta>0$ and
(\ref{eq:enum-zeros-poles})--(\ref{eq:bounded-below-neg})
yield $\mu_n-\lambda_n>0$ for all $n\in M$. Thus, taking into
account (\ref{eq:prod-posit-gen}), we obtain
\begin{equation}
  \label{eq:normalizing-posit-gen}
  \tau_n>0\,,\qquad\forall\, n\in M\,.
\end{equation}
Let us now define the function
\begin{equation}
  \label{eq:rho-thru-normalizing-def-gen}
  \rho(t):=\sum_{\lambda_k\le t}\frac{1}{\tau_k}\,,\quad
  t\in\mathbb{R}\,.
\end{equation}
Since (\ref{eq:normalizing-posit-gen}) holds, $\rho$ is a
monotone non-decreasing function and has an infinite number of
points of growth. Notice also that $\rho$ is right continuous.
Now, we want to show that for the measure corresponding to
$\rho$ all the moments are finite and
\begin{equation}
  \label{eq:measure-normal-gen}
  \int_{\mathbb{R}}d\rho(t)=1\,.
\end{equation}
The fact that the moments are finite follows directly from
condition \emph{\ref{finite-moments-sufficient}}). Indeed,
\begin{equation*}
  \int_{\mathbb{R}}t^md\rho(t) =
  \sum_{k\in M}\frac{\lambda_k^m}{\tau_k}\,.
\end{equation*}
We show next that (\ref{eq:measure-normal-gen}) holds true.
Given the sequences $\{\lambda_k\}_k$ and $\{\mu_k\}_k$
satisfying \emph{\ref{interlace-sufficient}}) and
\emph{\ref{sum-spectr-sufficient}}) we can define the function
  \begin{equation}
    \label{eq:m-tilde-def-gen}
    \widetilde{\mathfrak{m}}(\zeta):=
    \prod_{k\in M}
    \frac{\mu_k-\zeta}{\lambda_k-\zeta}\,.
  \end{equation}
Taking into account (\ref{eq:tau-def-1}), one obtains
  that
\begin{equation*}
  \res_{\zeta=\lambda_n}
  (\widetilde{\mathfrak{m}}(\zeta) -1)
  =-\frac{\Delta}{\tau_n}\,.
\end{equation*}
In view of \emph{\ref{sum-spectr-sufficient}}), we easily find
that
\begin{equation}
  \label{eq:m-1-to-0}
  \begin{split}
  \lim_{\substack{\zeta\to\infty \\ \im \zeta\ge\epsilon}}
  (\widetilde{\mathfrak{m}}(\zeta) -1) &=
  \lim_{\substack{\zeta\to\infty \\ \im \zeta\ge\epsilon}}
 \prod_{k\in M} \frac{\mu_k-\zeta}{\lambda_k-\zeta} -1
  =\\&=
  \lim_{\substack{\zeta\to\infty \\ \im \zeta\ge\epsilon}}
 \prod_{k\in M}\left(1+
   \frac{\mu_k-\lambda_k}{\lambda_k-\zeta}\right)
- 1=0\,.
  \end{split}
\end{equation}
Thus, on the basis of \v{C}ebotarev's theorem on the
representation of meromorphic Herglotz functions
\cite[Chap. VII, Sec.1 Theorem 2]{MR589888}, one obtains
\begin{equation}
  \label{eq:m-thru-normalizing-gen}
  \widetilde{\mathfrak{m}}(\zeta)  - 1
    = \sum_{k\in M}\frac{\Delta}{(\lambda_k-\zeta)\tau_k}\,.
\end{equation}
We now define the function $\widetilde{m}(\zeta):=
\frac{\widetilde{\mathfrak{m}}(\zeta)-1}{\Delta}$. Then,
(\ref{eq:m-thru-normalizing-gen}) yields
\begin{equation}
  \label{eq:m-normalizing-gen}
 \widetilde m(\zeta) =
 \sum_{k\in M}\frac{1}{\tau_k(\lambda_k-\zeta)}\,.
\end{equation}
We next show that
\begin{equation*}
  \lim_{\substack{\zeta\to\infty \\ \im \zeta\ge\epsilon}}
  \zeta\widetilde m(\zeta) = -1\,.
\end{equation*}
Indeed,
\begin{equation*}
  \begin{split}
  \frac{\widetilde{\mathfrak{m}}(\zeta)}{\Delta}&=
  \frac{1}{\Delta}\prod_{k\in M}
    \frac{\mu_k-\zeta}{\lambda_k-\zeta}=\\
    &=\frac{1}{\Delta}
    \exp\left\{\sum_{k\in M}\ln\left(
    \frac{\mu_k-\zeta}{\lambda_k-\zeta}\right)\right\}=\\
 &=\frac{1}{\Delta}
    \exp\left\{\sum_{k\in M}\ln\left( 1+
    \frac{\mu_k-\lambda_k}{\lambda_k-\zeta}\right)\right\}=\\
 &=\frac{1}{\Delta}
    \exp\left\{\sum_{k\in M}\sum_{p=1}^\infty(-1)^{p-1}\left(
    \frac{\mu_k-\lambda_k}{\lambda_k-\zeta}\right)^p\right\}\,.
\end{split}
\end{equation*}
Thus, as $\zeta\to\infty$ with $\im\zeta\ge\epsilon$
($\epsilon>0$),
\begin{equation*}
  \frac{\widetilde{\mathfrak{m}}(\zeta)}{\Delta}
  =\frac{1}{\Delta}+ \frac{1}{\Delta}
  \sum_{k\in M}\frac{\mu_k-\lambda_k}{\lambda_k-\zeta}
  + O(\zeta^{-2})\,.
\end{equation*}
Then,
\begin{equation*}
  \begin{split}
  \lim_{\substack{\zeta\to\infty \\ \im \zeta\ge\epsilon}}
  \zeta\widetilde m(\zeta)&=
\lim_{\substack{\zeta\to\infty \\ \im \zeta\ge\epsilon}}
  \zeta
  \frac{1}{\Delta}
\sum_{k\in M}\frac{\mu_k-\lambda_k}{\lambda_k-\zeta}=\\&=
-\frac{1}{\Delta}\sum_{k\in M}(\mu_k-\lambda_k)=-1\,.
\end{split}
\end{equation*}
Also, from
(\ref{eq:m-normalizing-gen}) one has
\begin{equation*}
  \lim_{\substack{\zeta\to\infty \\ \im \zeta\ge\epsilon}}
  \zeta\widetilde m(\zeta) =-\sum_{k\in M}\frac{1}{\tau_k}\,.
\end{equation*}
Therefore,
\begin{equation*}
  1=\sum_{k\in M}\frac{1}{\tau_k}=\int_\mathbb{R}d\rho(t)\,.
\end{equation*}
Having found a function $\rho$ with infinitely many growing
points and such that (\ref{eq:measure-normal-gen}) is
satisfied and all the moments exist, one can obtain, applying
the method of orthogonal polynomials (see Section 2), a
tridiagonal semi-infinite matrix. Let us denote by
$\widehat{J}$ the operator whose matrix representation is the
obtained matrix.  By what has been explained before, this
operator is closed and symmetric. Now, define
$h_2:=\Delta+h_1$ and $J:= \widehat{J}+ h_2
\langle\cdot,e_1\rangle\, e_1$.

If $\widehat{J}=\widehat{J}^*$, we know that $\rho(t)=\langle
E(t)_{ \widehat{J}}\, e_1,e_1\rangle$, where
$E_{\widehat{J}}(t)$ is the spectral decomposition of the
self-adjoint Jacobi operator $\widehat{J}$. Then, obviously,
$\widehat{J}= J_{h_2}$.

If $\widehat{J}\ne \widehat{J}^*$, the Stieltjes transform of
$\rho$ is the Weyl $m$-function, we denote it by $w(\zeta)$,
of some self-adjoint extension of $\widehat{J}$ that we denote
by $\widetilde{J}$. This is true because of the density of
polynomials in $L_2(\mathbb{R},d\rho)$. Indeed,
\emph{\ref{density-poly-sufficient}}) means that the
polynomials are dense in $L_2(\mathbb{R},d\rho)$. Thus,
$w(\zeta)$ lies on the Weyl circle, and then, it is the Weyl
$m$-function of some self-adjoint extension of $\widehat{J}$
\cite{MR0184042}, \cite[Proposition
4.15]{MR1627806}. Therefore, $ \widetilde{J}+ h_2
\langle\cdot,e_1\rangle\, e_1$ is a self-adjoint extension of
$J$ and hence, $ \widetilde{J}+ h_2 \langle\cdot,e_1\rangle\,
e_1= J(g)$ for some unique $g \in \mathbb R \cup \{\infty\}$.
Furthermore, we obviously have that, $\widetilde{J}=
J_{h_2}(g)$ and $w(\zeta)=m_{h_2}(\zeta,g)$. We uniquely
reconstruct $m(\zeta,g)$ from $m_{h_2}(\zeta,g)$ using
(\ref{eq:m-two-boundaries}) and then, we uniquely reconstruct
$g$ as explained in Section 2.

Notice that we have
\begin{equation*}
   m_{h_2}(\zeta,g)=\int_\mathbb{R}\frac{d\rho(t)}{t-\zeta}
   =\widetilde{m}(\zeta)\,.
\end{equation*}

It remains to show that $\sigma(J_{h_2}(g))=\{\lambda_k\}_k$
and $\sigma(J_{h_1}(g))=\{\mu_k\}_k$. To this end consider the
function $\mathfrak{m}(\zeta,g)$ for the pair $J_{h_2}$ and
$J_{h_1}$:
\begin{equation*}
  \mathfrak{m}(\zeta,g)=
  \frac{m_{h_2}(\zeta,g)}{m_{h_1}(\zeta,g)}\,,
  \qquad\zeta\in\mathbb{C}\setminus\mathbb{R}\,.
\end{equation*}
Let the sequence $\{\gamma_k\}_k$ denote the spectrum of
$J_{h_1}$. Then, arguing as in the proof of (\ref{eq:short})
we obtain that
\begin{equation*}
  \mathfrak{m}(\zeta,g)=\prod_{k\in M}
\frac{\gamma_k-\zeta}{\lambda_k-\zeta}\,.
\end{equation*}
Since we have already proven that
\emph{\ref{interlace-sufficient}}) and
\emph{\ref{sum-spectr-sufficient}}) are necessary conditions,
we have that
\begin{equation*}
   \sum_{k\in M}(\gamma_k-\lambda_k)=\Delta<\infty\,.
\end{equation*}
Then, as in the proof of (\ref{eq:m-1-to-0}), it follows that
\begin{equation*}
  \lim_{\substack{\zeta\to\infty \\ \im \zeta\ge\epsilon}}
  (\mathfrak{m}(\zeta) -1)=0\,.
\end{equation*}
Hence by \v{C}ebotarev's theorem
\cite[Chap. VII, Sec.1 Theorem 2]{MR589888},
\begin{equation*}
  \mathfrak{m}(\zeta,g)=1+\sum_{k\in M}
  \frac{h_2-h_1}{(\lambda_k-\zeta)\alpha_k(h_2,g)}\,,
\end{equation*}
where we compute the residues of $\mathfrak{m}(\zeta)$ as
in (\ref{eq:res-m-computation}). Thus, since
$\alpha_k(h_2,g)=\tau_k$, $\forall k\in M$,
\begin{equation*}
  \mathfrak{m}(\zeta,g)=
  1+\sum_{k\in M}
  \frac{\Delta}{(\lambda_k-\zeta)\tau_k}=
  \widetilde{\mathfrak{m}}(\zeta,g)\,.
\end{equation*}
But $\{\lambda_k\}_k$ and $\{\mu_k\}_k$ are the poles and
zeros of $\widetilde{\mathfrak{m}}(\zeta,g)$ and then, the
eigenvalues of $J_{h_2}(g)$ and
$J_{h_1}(g)$, respectively.\\
\end{proof}
\begin{remark}
  \label{rem:1-1}
  We draw the reader's attention to the fact that the matrix
  associated with the function $\rho$, constructed in the
  proof of the previous theorem, may have deficiency indices
  $(1,1)$ \cite{MR0184042,MR1627806,MR1711536}.

  If we drop the condition of the density of polynomials in
  $L_2(\mathbb{R},d\rho)$ and our reconstruction method yields
  a nonself-adjoint operator $J$, then the sequences
  $\{\lambda_k\}_k$ and $\{\mu_k\}_k$ correspond to the
  spectra of some generalized self-adjoint extensions of
  $J_{h_2}$ and $J_{h_1}$, respectively (see
  \cite{MR1627806}). The generalized extensions of symmetric
  operators, which are not von Neumann extensions, were first
  introduced by Naimark (see Appendix I in \cite{MR1255973} on
  Naimark's theory).
\end{remark}

In \cite{MR0221315} the case of Jacobi operators bounded from
below is considered.  A uniqueness result is proven, and some
sufficient conditions for a pair of sequences to be the
spectra of a Jacobi operator with different boundary
conditions are given.

\section{Dirichlet-Neumann conditions}\sss
\label{sec:d-n-cond}

\subsection{Recovering the matrix from two spectra}
\label{sec:recov-matr-from}

In this section we shall consider the pair of Jacobi operators
$J_0(g)=J(g)$ and $J_\infty(g)$.  Here, as before, we keep the
convention of writing $J(g)$ even if $J=J^*$. The matrix
representation of $J_\infty(g)$ corresponds to the matrix
representation of $J(g)$ with the first column and row
removed.  From the Ricatti equation (\ref{eq:ricatti}), taking
into account that $m^{(0)}(\zeta,g)=m(\zeta,g)$ and
$m^{(1)}(\zeta,g)=m_\infty(\zeta,g)$, we have
\begin{equation}
  \label{eq:relation-weyl-functions}
   m_\infty(\zeta,g)=-\frac{1}{b_1^2}\left((\zeta-q_1)
  + \frac{1}{m(\zeta,g)}\right)\,.
\end{equation}

As before, we assume that the spectrum of $J=J(g)$ is
discrete.

If $m(\zeta,g)$ is a meromorphic function, then, by
(\ref{eq:relation-weyl-functions}), $m_\infty(\zeta,g)$ is
also meromorphic and the spectrum of $J_\infty(g)$ is
discrete. The poles of $m(\zeta,g)$ are the eigenvalues of
$J(g)$, while the zeros of $m(\zeta,g)$ are the eigenvalues
of $J_\infty(g)$.  Since $m(\zeta,g)$ is always a Herglotz
function, under our assumption on the discreteness of
$\sigma(J(g))$, $m(\zeta,g)$ is a meromorphic Herglotz
function.  This implies that $\sigma(J(g))$ and
$\sigma(J_\infty(g))$ are interlaced, that is, between two
successive eigenvalues of one operator there is exactly one
eigenvalue of the other.

Let the sequence $\{\lambda_k\}_k$ denote the eigenvalues of
$J(g)$ (the poles of $m(\zeta,g)$). Furthermore, $\{\mu_k\}_k$
will stand for the eigenvalues of $J_\infty(g)$ (the zeros of
$m(\zeta,g)$). It is worth remarking that, in contrast to the
case of boundary conditions being rank one perturbations with
finite coupling constant, here our convention for enumerating
the sequences $\{\lambda_k\}_k$ and $\{\mu_k\}_k$ does not
work in the case when $J(g)$ is semi-bounded from above.
Indeed, it follows from the mini-max principle \cite{rs} that
if $J(g)$ is bounded from below, the smallest of all poles is
less than the smallest of all zeros of $m(\zeta,g)$, and if
$J(g)$ is bounded from above, the min-max principle applied to
$-J(g)$ implies that the greatest of all zeros is less than
the greatest of all poles of $m(\zeta,g)$.

So let us consider first the case when $J(g)$ is not
semi-bounded or semi-bounded from below and enumerate the
sequences $\{\lambda_k\}_k$ and $\{\mu_k\}_k$ by
(\ref{eq:enum-zeros-poles}), (\ref{eq:unbounded-case}),
(\ref{eq:bounded-below-0}), and (\ref{eq:bounded-below-neg}). Then, by
the same theorem we used to obtain
(\ref{eq:levin-herglotz-gen}) \cite{MR589888}, $m(\zeta,g)$
can be written as follows
\begin{equation}
  \label{eq:levin-herglotz}
  m(\zeta,g) = C \frac{\zeta-\mu_0}{\zeta-\lambda_0}
  \sideset{}{'}\prod_{k\in M} \left(1-\frac{\zeta}{\mu_k}\right)
  \left(1-\frac{\zeta}{\lambda_k}\right)^{-1}\,,\qquad C>0\,,
\end{equation}
where, as before, the prime in the infinite product means that
it does not include the factor $k=0$.

If $J(g)$ is bounded from above, then we are still able to use
(\ref{eq:enum-zeros-poles}), (\ref{eq:bounded-above-positive}) and
(\ref{eq:bounded-above-0}) for enumerating the zeros and poles
of the meromorphic Herglotz function $-\frac{1}{m(\zeta,g)}$.
Thus,
\begin{equation}
  \label{eq:levin-herglotz-inv}
  -\frac{1}{m(\zeta,g)}=\widetilde{C}
  \frac{\zeta-\lambda_0}{\zeta-\mu_0}
  \sideset{}{'}\prod_{k\in M} \left(1-\frac{\zeta}{\lambda_k}\right)
  \left(1-\frac{\zeta}{\mu_k}\right)^{-1}\,,\qquad
  \widetilde C>0\,.
\end{equation}
Notice that, since we have enumerated zeros and poles of
$-\frac{1}{m(\zeta,g)}$ by our convention, we have now
\begin{equation}
  \label{eq:enum-zeros-poles-alt}
   \forall k\in M\,,\quad\mu_k<\lambda_k<\mu_{k+1}\,,
\end{equation}
and
\begin{enumerate}[\ a)]
\item
if $ 0 < \sup_k\{\mu_k\}_k< \infty$,
\begin{equation}
  \label{eq:bounded-above-pos-alt}
  M:= \{k\}_{k=-\infty}^{k_{\max}}\,,
  \ (k_{\max}\ge 1)\quad\text{requiring}\quad
\lambda_{-1} < 0 < \mu_1\,,
\end{equation}
\item
if $\sup_k\{\mu_k\}_k\le 0$,
\begin{equation}
  \label{eq:bounded-above-0-alt}
  M:=\{k\}_{k=-\infty}^0\,.
\end{equation}
\end{enumerate}
Here again $\lambda_0$ or $\mu_0$ are the only ones allowed to be
zero.

Equations (\ref{eq:levin-herglotz}) and
(\ref{eq:levin-herglotz-inv}) can be written in one formula
\begin{equation}
  \label{eq:levin-herglotz-yes-not}
   m(\zeta,g) = K \frac{\zeta-\mu_0}{\zeta-\lambda_0}
  \sideset{}{'}\prod_{k\in M} \left(1-\frac{\zeta}{\mu_k}\right)
  \left(1-\frac{\zeta}{\lambda_k}\right)^{-1}\,,
\end{equation}
where, if $J(g)$ is not semi-bounded from above, $K=C$ and
$\{\lambda_k\}_k$ and $\{\mu_k\}_k$ are enumerated by
(\ref{eq:enum-zeros-poles}), (\ref{eq:unbounded-case}),
(\ref{eq:bounded-below-0}), and (\ref{eq:bounded-below-neg}),
while $K=-\widetilde{C}^{-1}$ and $\{\lambda_k\}_k$ and
$\{\mu_k\}_k$ are enumerated by
(\ref{eq:enum-zeros-poles-alt})--(\ref{eq:bounded-above-0-alt})
if $J(g)$ is semi-bounded from above.

We give now, for the reader's convenience, a simple proof of a
theorem that was proven by Fu and Hochstadt \cite{MR49:9676}
for regular Jacobi operators (a regular Jacobi matrix is
defined in \cite{MR49:9676}), and by Teschl \cite{MR1643529}
in the general case.

\begin{theorem}(Fu and Hochstadt, Teschl)
  \label{uniq.2}
  Consider the Jacobi operator $J(g)$ with discrete spectrum.
  The sequences $\{\lambda_k\}_k=\sigma(J(g))$ and
  $\{\mu_k\}_k=\sigma(J_\infty(g))$ uniquely determine the
  operator $J$ and, if $J\ne J^*$, the boundary condition, $g$,
  at infinity.
\end{theorem}
\begin{proof}
  From (\ref{eq:m-asympt}) we know that
  \begin{equation*}
    \lim_{\substack{\zeta\to\infty \\ \im \zeta\ge\epsilon}}
    \zeta m(\zeta,g)=-1\,,\qquad \epsilon>0\,.
  \end{equation*}
  Then, if $J(g)$ is not semi-bounded from above,
  (\ref{eq:levin-herglotz}) yields
\begin{equation}
  \label{eq:c-deter-2-spect}
  C^{-1}=-\lim_{\substack{\zeta\to\infty \\ \im \zeta\ge\epsilon}}\zeta
  \sideset{}{'}\prod_{k\in M}
  \left(1-\frac{\zeta}{\mu_k}\right)
  \left(1-\frac{\zeta}{\lambda_k}\right)^{-1}\,,
  \qquad \epsilon>0\,,
\end{equation}
where $\{\lambda_k\}_k$ and $\{\mu_k\}_k$ are enumerated by
(\ref{eq:enum-zeros-poles}), (\ref{eq:unbounded-case}),
(\ref{eq:bounded-below-0}), and (\ref{eq:bounded-below-neg}).
On the other hand, in the semi-bounded from above case
(\ref{eq:levin-herglotz-inv}) implies
\begin{equation}
  \label{eq:tilde-c-deter-2-spect}
  \widetilde{C}^{-1}=
  \lim_{\substack{\zeta\to\infty \\ \im \zeta\ge\epsilon}}
  \frac{1}{\zeta}
  \sideset{}{'}\prod_{k\in M}
  \left(1-\frac{\zeta}{\lambda_k}\right)
  \left(1-\frac{\zeta}{\mu_k}\right)^{-1}\,,
  \qquad \epsilon>0\,.
\end{equation}
where $\{\lambda_k\}_k$ and $\{\mu_k\}_k$ are enumerated by
(\ref{eq:enum-zeros-poles-alt})--(\ref{eq:bounded-above-0-alt}).
Thus, in any case, one can find $K$, the constant in
(\ref{eq:levin-herglotz-yes-not}), from the sequences
$\{\lambda_k\}_k$ and $\{\mu_k\}_k$.  Therefore, the spectra
$\sigma(J(g))$ and $\sigma(J_\infty(g))$ uniquely determine
$m(\zeta,g)$. Having found $m(\zeta,g)$ we can, using the
methods introduced in Section \ref{sec:preliminaries},
determine $J$ and, in the case when $J\ne J^*$, also find
uniquely the boundary condition at infinity, $g$.\\
\end{proof}
\begin{remark}
\label{rem:K-expression}
  It turns out that, by (\ref{eq:c-deter-2-spect}) and
  (\ref{eq:tilde-c-deter-2-spect}), $K$ can be written as
  \begin{equation}
    \label{eq:K-gen-expression}
    K^{-1}=-\lim_{\substack{\zeta\to\infty \\ \im \zeta\ge\epsilon}}\zeta
  \sideset{}{'}\prod_{k\in M}
  \left(1-\frac{\zeta}{\mu_k}\right)
  \left(1-\frac{\zeta}{\lambda_k}\right)^{-1}\,,
  \qquad \epsilon>0\,,
  \end{equation}
  where the sequences $\{\lambda_k\}_k$ and $\{\mu_k\}_k$ have
  been enumerated by (\ref{eq:enum-zeros-poles}),
  (\ref{eq:unbounded-case}), (\ref{eq:bounded-below-0}), and
  (\ref{eq:bounded-below-neg}), when $J(g)$ is not
  semi-bounded from above and by
  (\ref{eq:enum-zeros-poles-alt})--(\ref{eq:bounded-above-0-alt}),
  otherwise.
\end{remark}

In what follows the Weyl $m$-function will be written through
(\ref{eq:levin-herglotz-yes-not}) with $K$ given by
(\ref{eq:K-gen-expression}). From
(\ref{eq:levin-herglotz-yes-not}) one can obtain
straightforward formulae for the normalizing constants
(\ref{eq:def-normalizing}) in terms of the sequences
$\{\lambda_k\}_k$ and $\{\mu_k\}_k$. Indeed, when $n\ne 0$
\begin{equation*}
\begin{split}
  \lim_{\zeta\to\lambda_n}(\lambda_n-\zeta)m(\zeta,g)&=
\lim_{\zeta\to\lambda_n}(\lambda_n-\zeta)
K \frac{\zeta-\mu_0}{\zeta-\lambda_0}
  \sideset{}{'}\prod_{k\in M}\frac{
  1-\frac{\zeta}{\mu_k}}
  {1-\frac{\zeta}{\lambda_k}}=\\[3mm]
  &=K\frac{\lambda_n}{\mu_n}(\mu_n-\lambda_n)
  \frac{\lambda_n-\mu_0}{\lambda_n-\lambda_0}
\sideset{}{'}\prod_{\substack{k\in M\\k\ne n}}\frac{
  1-\frac{\lambda_n}{\mu_k}}
  {1-\frac{\lambda_n}{\lambda_k}}\ .
\end{split}
\end{equation*}
Formulae (\ref{eq:normalizing-const-formula}) and
(\ref{eq:K-gen-expression}) then give
\begin{equation}
  \label{eq:n-normalizing-spect}
  \alpha_n^{-1}=-\frac{
  \frac{\lambda_n}{\mu_n}(\mu_n-\lambda_n)\frac{\lambda_n-\mu_0}
    {\lambda_n-\lambda_0}
  \sideset{}{'}\prod_{\substack{k\in M\\k\ne n}}
 \left(1-\frac{\lambda_n}{\mu_k}\right)
  \left(1-\frac{\lambda_n}{\lambda_k}\right)^{-1}}
  {\lim_{\substack{\zeta\to\infty \\ \im \zeta\ge\epsilon}}\zeta
  \sideset{}{'}\prod_{k\in M}
  \left(1-\frac{\zeta}{\mu_k}\right)
  \left(1-\frac{\zeta}{\lambda_k}\right)^{-1}}\,,
  \quad n\ne 0\,.
\end{equation}
Analogously,
\begin{equation}
 \label{eq:0-normalizing-spect}
  \alpha_0^{-1}= -\frac{(\mu_0-\lambda_0)
  \sideset{}{'}\prod_{k\in M}
  \left(1-\frac{\lambda_0}{\mu_k}\right)
  \left(1-\frac{\lambda_0}{\lambda_k}\right)^{-1}}
 {\lim_{\substack{\zeta\to\infty \\ \im \zeta\ge\epsilon}}\zeta
  \sideset{}{'}\prod_{k\in M}
  \left(1-\frac{\zeta}{\mu_k}\right)
  \left(1-\frac{\zeta}{\lambda_k}\right)^{-1}}
  \,.
\end{equation}

\subsection{Necessary and Sufficient Conditions }
\label{sec:suff-cond-two}
The following result establishes necessary and sufficient
conditions for two given sequences of real numbers to be the
spectra of $J(g)$ and $J_\infty(g)$.
\begin{theorem}
  \label{thm:dn-suff-cond-two-gen}
  Given two infinite sequences of real numbers
  $\{\lambda_k\}_k$ and $\{\mu_k\}_k$ without finite points of
  accumulation, there is a unique operator $J(g)$, and if
  $J\ne J^\ast$ also a unique $ g \in \mathbb R \cup
  \{+\infty\}$, such that $\{\lambda_k\}_k=\sigma(J(g))$ and
  $\{\mu_k\}_k=\sigma(J_\infty(g))$ if and only if the
  following conditions are satisfied.
  \begin{enumerate}[\ a)]
  \item $\{\lambda_k\}_k$ and $\{\mu_k\}_k$ interlace and, if
    $\{\lambda_k\}_k$ is bounded from below,
    $\min_k\{\mu_k\}_k>\min_k\{\lambda_k\}_k$, if
    $\{\lambda_k\}_k$ is bounded from above,
    $\max_k\{\lambda_k\}_k>\max_k\{\mu_k\}_k$. So we use below
    the convention (\ref{eq:enum-zeros-poles}),
    (\ref{eq:unbounded-case}), (\ref{eq:bounded-below-0}), and
    (\ref{eq:bounded-below-neg}) for enumerating the sequences
    when $J(g)$ is not semi-bounded from above, and
    (\ref{eq:enum-zeros-poles-alt})--(\ref{eq:bounded-above-0-alt})
    otherwise.\label{dn-interlace-suff}

 By condition \ref{dn-interlace-suff}) the product
    \begin{equation*}
      \sideset{}{'}\prod_{k\in M}
  \left(1-\frac{\zeta}{\mu_k}\right)
  \left(1-\frac{\zeta}{\lambda_k}\right)^{-1}
    \end{equation*}
    converges uniformly on compact subsets of $\mathbb{C}$
    (see the proof below and \cite[Chap. 7, Sec.1]{MR589888}).
  \item  The limit
  \begin{equation}
    \label{eq:limit-over-imaginary}
    \lim_{\substack{\xi\to\infty \\ \xi\in\mathbb{R}}}\I\xi
  \sideset{}{'}\prod_{k\in M}
  \left(1-\frac{\I\xi}{\mu_k}\right)
  \left(1-\frac{\I\xi}{\lambda_k}\right)^{-1}
  \end{equation}
  is finite and negative when the sequences $\{\lambda_k\}_k$
  and $\{\mu_k\}_k$ are not bounded from above, and it is
  finite and positive otherwise.\label{dn-limit-sufficient}
\item Let $\{\tau_n\}_{n\in M}$ be defined by
  \begin{equation*}
    \tau_n^{-1}=-\frac{
  \frac{\lambda_n}{\mu_n}(\mu_n-\lambda_n)\frac{\lambda_n-\mu_0}
    {\lambda_n-\lambda_0}
  \sideset{}{'}\prod_{\substack{k\in M\\k\ne n}}
 \left(1-\frac{\lambda_n}{\mu_k}\right)
  \left(1-\frac{\lambda_n}{\lambda_k}\right)^{-1}}
  {\lim_{\substack{\xi\to\infty \\ \xi\in\mathbb{R}}}\I\xi
  \sideset{}{'}\prod_{k\in M}
  \left(1-\frac{\I\xi}{\mu_k}\right)
  \left(1-\frac{\I\xi}{\lambda_k}\right)^{-1}}\,,
  \quad  n\in M\,,\ n\ne 0\,,
    \end{equation*}
and
\begin{equation*}
  \tau_0^{-1}= -\frac{(\mu_0-\lambda_0)
  \sideset{}{'}\prod_{k\in M}
  \left(1-\frac{\lambda_0}{\mu_k}\right)
  \left(1-\frac{\lambda_0}{\lambda_k}\right)^{-1}}
  {\lim_{\substack{\xi\to\infty \\ \xi\in\mathbb{R}}}\I\xi
  \sideset{}{'}\prod_{k\in M}
  \left(1-\frac{\I\xi}{\mu_k}\right)
  \left(1-\frac{\I\xi}{\lambda_k}\right)^{-1}}.
\end{equation*}
The sequence $\{\tau_n\}_{n\in M}$ is such that, for
$m=0,1,2\dots$, the series
\begin{equation*}
    \sum_{k\in M}\frac{\lambda_k^{2m}}{\tau_k}
    \qquad\text{converges}.
\end{equation*}\label{dn-finite-moments-sufficient}
\item  If a sequence of complex numbers $\{\beta_k\}_{k\in M}$,
  is such that the series
  \begin{equation*}
    \sum_{k\in
      M}\frac{\abs{\beta_k}^2}{\tau_k}
\quad\text{converges}
  \end{equation*}
and, for $m=0,1,2,\dots$,
\begin{equation*}
  \sum_{k\in
    M}\frac{\beta_k\lambda_k^m}{\tau_k}=0\,,
\end{equation*}
then $\beta_k=0$ for all $k\in M$.\label{dn-density-sufficient}
  \end{enumerate}
\end{theorem}
\begin{proof}
  We begin the proof by showing that the sequences
  $\sigma(J(g))=\{\lambda_k\}_k$ and
  $\sigma(J_\infty(g))=\{\mu_k\}_k$ satisfy
  \emph{\ref{dn-interlace-suff}}),
  \emph{\ref{dn-limit-sufficient}}),
  \emph{\ref{dn-finite-moments-sufficient}}), and
  \emph{\ref{dn-density-sufficient}}). Since the Weyl $m$
  function is Herglotz, the eigenvalues of $J(g)$ and
  $J_\infty(g)$ interlace as indicated in
  \emph{\ref{dn-interlace-suff}}). To prove that
  \emph{\ref{dn-limit-sufficient}}) holds, consider first the
  case when $J(g)$ is not semi-bounded or only bounded from
  below, then (\ref{eq:c-deter-2-spect}) yields
  \emph{\ref{dn-limit-sufficient}}). If $J(g)$ is semi-bounded
  from above, (\ref{eq:tilde-c-deter-2-spect}) implies
  \emph{\ref{dn-limit-sufficient}}).

  On the basis of (\ref{eq:n-normalizing-spect}) and
  (\ref{eq:0-normalizing-spect}), $\tau_n$ coincides with the
  normalizing constant $\alpha_n$ for all $n\in M$. Hence the
  spectral function $\rho$ of the self-adjoint extension
  $J(g)$ is given by the expression
  $\rho(t)=\sum_{\lambda_k\le t}\tau_k^{-1}$. Thus
  \emph{\ref{dn-finite-moments-sufficient}}) follows from the
  fact that all the moments of $\rho$ are finite
  \cite{MR0184042,MR1627806}. Similarly,
  \emph{\ref{dn-density-sufficient}}) stems from the density
  of polynomials in $L_2(\mathbb{R},d\rho)$, which takes place
  since $\rho$ is $N$-extremal \cite{MR0184042},
  \cite[Proposition 4.15]{MR1627806}.

  Let us now suppose that we are given two real sequences
  $\{\lambda_k\}_k$ and $\{\mu_k\}_k$ that satisfy
  \emph{\ref{dn-interlace-suff}}).  It can be shown
  that
\begin{equation}
  \label{eq:prod-posit}
  0<\sideset{}{'}\prod_{\substack{k\in M\\k\ne n}}
 \left(1-\frac{\lambda_n}{\mu_k}\right)
  \left(1-\frac{\lambda_n}{\lambda_k}\right)^{-1}<\infty\,.
\end{equation}
Indeed, the convergence of the infinite product follows from
\emph{\ref{dn-interlace-suff}}) and is part of the Theorem 1
in \cite[Chap. 7, Sec.1]{MR589888} used to obtain
(\ref{eq:levin-herglotz-gen}). We give here, for the reader's
convenience, some details. The product in
(\ref{eq:prod-posit}) converges if and only if
\begin{equation*}
    \sideset{}'\sum_{\substack{k\in M\\k\ne n}}\left\{
  \left(1-\frac{\lambda_n}{\mu_k}\right)
\left(1-\frac{\lambda_n}{\lambda_k}\right)^{-1}
-1\right\}=\lambda_n\sideset{}'\sum_{\substack{k\in M\\k\ne n}}
    \left(\frac{1}{\lambda_k}-\frac{1}{\mu_k}\right)
\left(1-\frac{\lambda_n}{\lambda_k}\right)^{-1}<\infty\,,
\end{equation*}
where prime means that the summand $k=0$ is excluded.
Thus, we have to prove that
  \begin{equation*}
    \sideset{}'\sum_{k\in M}
    \left(\frac{1}{\lambda_k}-\frac{1}{\mu_k}\right)<\infty\,.
  \end{equation*}
It will suffice to consider that in
 \emph{\ref{dn-interlace-suff}}) the sequences are ordered by
 (\ref{eq:enum-zeros-poles}) with $M$ given by
 (\ref{eq:unbounded-case}). For any $k\in\mathbb{N}$,
 (\ref{eq:enum-zeros-poles}) implies
\begin{equation*}
   0< \left(\frac{1}{\lambda_k}-\frac{1}{\mu_k}\right)<
    \left(\frac{1}{\lambda_k}-\frac{1}{\lambda_{k+1}}\right)\,,
    \quad\forall k\in\mathbb{N}\,.
\end{equation*}
Clearly,
$\sum_{k\in\mathbb{N}}\left(\frac{1}{\lambda_k}
-\frac{1}{\lambda_{k+1}}\right)$ is convergent. Analogously, it
can be proven that
\begin{equation*}
  \sum_{k\in\mathbb{N}}\left(\frac{1}{\lambda_{-k}}
-\frac{1}{\mu_{-k}}\right)<\infty\,.
\end{equation*}
Having established the convergence of the the product in
(\ref{eq:prod-posit}), its positivity follows easily.

We have, therefore, a sequence of real numbers $\{\tau_k\}_{k\in
  M}$ and let us now show that $\tau_n>0$, $\forall n\in M$.
First notice that (\ref{eq:enum-zeros-poles}),
(\ref{eq:unbounded-case}), (\ref{eq:bounded-below-0}), and
(\ref{eq:bounded-below-neg}), yield
\begin{equation*}
  \frac{\lambda_n}{\mu_n}(\mu_n-\lambda_n)\frac{\lambda_n-\mu_0}
    {\lambda_n-\lambda_0}>0\quad
    (n\ne 0)\quad\text{ and }\quad\mu_0-\lambda_0>0\,.
\end{equation*}
On the other hand
(\ref{eq:enum-zeros-poles-alt})--(\ref{eq:bounded-above-0-alt})
imply
\begin{equation*}
  \frac{\lambda_n}{\mu_n}(\mu_n-\lambda_n)\frac{\lambda_n-\mu_0}
    {\lambda_n-\lambda_0}<0\quad
    (n\ne 0)
\quad\text{ and }\quad\mu_0-\lambda_0<0\,.
\end{equation*}
From these last inequalities, taking into account
(\ref{eq:prod-posit}) and condition
\emph{\ref{dn-limit-sufficient}}) we obtain
\begin{equation}
  \label{eq:normalizing-posit}
  \tau_n>0\,,\qquad\forall\, n\in M\,.
\end{equation}
Let us now define the function
\begin{equation}
  \label{eq:rho-thru-normalizing-def}
  \rho(t):=\sum_{\lambda_k\le t}\frac{1}{\tau_k}\,,\qquad\forall
  t\in\mathbb{R}\,.
\end{equation}
In view of (\ref{eq:normalizing-posit}), $\rho$ is a monotone
non-decreasing function and has an infinite number of points
of growth. Now, we want to show that, for the measure
corresponding to $\rho$, all the moments are finite and
\begin{equation}
  \label{eq:measure-normal}
  \int_{\mathbb{R}}d\rho(t)=1\,.
\end{equation}
The fact that the moments are finite follows directly from
condition \emph{\ref{dn-finite-moments-sufficient}}). Indeed,
\begin{equation*}
  \int_{\mathbb{R}}t^md\rho(t) =
  \sum_{k\in M}\frac{\lambda_k^m}{\tau_k}\,.
\end{equation*}
We show next that (\ref{eq:measure-normal}) is true. Given
the sequences $\{\lambda_k\}_k$ and $\{\mu_k\}_k$ satisfying
\emph{\ref{dn-interlace-suff}}) and
\emph{\ref{dn-limit-sufficient}}), we can define the function
\begin{equation}
    \label{eq:m-tilde-def}
    \widetilde{m}(\zeta):=
    -\frac{\frac{\zeta-\mu_0}{\zeta-\lambda_0}
  \sideset{}{'}\prod_{k\in M} \left(1-\frac{\zeta}{\mu_k}\right)
  \left(1-\frac{\zeta}{\lambda_k}\right)^{-1}}
  {\lim_{\substack{\xi\to\infty \\ \xi\in\mathbb{R}}}\I\xi
  \sideset{}{'}\prod_{k\in M}
  \left(1-\frac{\I\xi}{\mu_k}\right)
  \left(1-\frac{\I\xi}{\lambda_k}\right)^{-1}}\,.
\end{equation}
Now, arguing as in the proof of (\ref{eq:n-normalizing-spect})
and (\ref{eq:0-normalizing-spect}), we obtain
\begin{equation*}
  \res_{\zeta=\lambda_n}\widetilde m(\zeta)=-\tau_n^{-1}\,.
\end{equation*}
On the other hand,
\begin{equation*}
  \lim_{\substack{\xi\to\infty \\ \xi\in\mathbb{R}}}
  \widetilde m(\I\xi) =
  -\lim_{\substack{\xi\to\infty \\ \xi\in\mathbb{R}}}
  \frac{\sideset{}{'}\prod_{k\in M} \left(1-\frac{\I\xi}{\mu_k}\right)
  \left(1-\frac{\I\xi}{\lambda_k}\right)^{-1}}
  {\I\xi
  \sideset{}{'}\prod_{k\in M}
  \left(1-\frac{\I\xi}{\mu_k}\right)
  \left(1-\frac{\I\xi}{\lambda_k}\right)^{-1}}=0\,.
\end{equation*}
Thus, using again \v{C}ebotarev's theorem \cite{MR589888} we
find that
\begin{equation}
  \label{eq:m-thru-normalizing}
  \widetilde m(\zeta) =
\sum_{k\in M}\frac{1}{\tau_k(\lambda_k-\zeta)}\,.
\end{equation}
It follows from (\ref{eq:m-tilde-def}) that
\begin{equation*}
  \lim_{\substack{\xi\to\infty \\ \xi\in\mathbb{R}}}
  \I\xi\widetilde m(\I\xi) =
  -\lim_{\substack{\xi\to\infty \\\xi\in\mathbb{R} }}
  \frac{\I\xi\sideset{}{'}\prod_{k\in M}
\left(1-\frac{\I\xi}{\mu_k}\right)
  \left(1-\frac{\I\xi}{\lambda_k}\right)^{-1}}
  {\I\xi
  \sideset{}{'}\prod_{k\in M}
  \left(1-\frac{\I\xi}{\mu_k}\right)
  \left(1-\frac{\I\xi}{\lambda_k}\right)^{-1}}=-1\,.
\end{equation*}
Also from (\ref{eq:m-thru-normalizing}) one has
\begin{equation*}
  \lim_{\substack{\xi\to\infty \\ \xi\in\mathbb{R}}}
  \I\xi\widetilde m(\I\xi) =-\sum_{k\in M}\frac{1}{\tau_k}\,.
\end{equation*}
Therefore,
\begin{equation*}
  1=\sum_{k\in M}\frac{1}{\tau_k}=\int_\mathbb{R}d\rho(t)\,.
\end{equation*}
We have found a function $\rho(t)$ with infinitely many
growing points, such that all the moments exist for the
corresponding measure and (\ref{eq:measure-normal}) holds.
Therefore one can obtain, applying the method of orthogonal
polynomials (see Section 2), a tridiagonal semi-infinite
matrix.  Let us denote by $J$ the operator whose matrix
representation is the obtained matrix.  As was mentioned
before, $J$ is symmetric and closed. Now, if $J=J^*$, we know
that $\rho(t)=\langle E_{J}(t)e_1,e_1\rangle$, where
$E_{J}(t)$ is the spectral decomposition of the self-adjoint
Jacobi operator $J$. If $J\ne J^*$, then the Stieltjes
transform of $\rho(t)$ is the Weyl $m$-function $m(\zeta,g)$
of some self-adjoint extension of $J$ with boundary conditions
at infinity given by $g$, that is,
\begin{equation*}
   m(\zeta,g)=\int_\mathbb{R}\frac{d\rho(t)}{t-\zeta}\,.
\end{equation*}
This last assertion is true because of the density of
polynomials in $L_2(\mathbb{R},d\rho)$, which follows from
\emph{\ref{dn-density-sufficient}}). Hence $\rho$ is
$N$-extremal \cite{MR0184042}. This  implies that
$m(\zeta,g)$ lies on the Weyl circle, and then it is the Weyl
$m$-function of some self-adjoint extension $J(g)$
\cite{MR0184042}, \cite{MR1627806}.

It remains to show that
$\sigma(J(g))=\{\lambda_k\}_k$ and
$\sigma(J_\infty(g))=\{\mu_k\}_k$.

So we start from (\ref{eq:rho-thru-normalizing-def}) and find
the Weyl $m$-function of $J(g)$ using
(\ref{eq:m-thru-normalizing})
\begin{equation*}
  m(\zeta,g)=\int_\mathbb{R}\frac{d\rho(t)}{t-\zeta}=
  \sum_{k\in M}\frac{1}{\tau_k(\lambda_k-\zeta)}=
  \widetilde m(\zeta)\,.
\end{equation*}
But $\{\lambda_k\}_k$ and $\{\mu_k\}_k$ are the poles and
zeros of $\widetilde m$ and then the eigenvalues of
$J(g)$ and $J_\infty(g)$,
respectively.\\
\end{proof}

For Jacobi operators semi-bounded from below, necessary and
sufficient conditions are given in \cite{MR499269}.  Note that
Remark \ref{rem:1-1} can also be made here.

It is worth mentioning that, from (\ref{eq:c-deter-2-spect})
and (\ref{eq:tilde-c-deter-2-spect}), it follows that, when
\emph{\ref{dn-limit-sufficient}}) is seen as a necessary
condition, one could write
\begin{equation*}
  \lim_{\substack{\zeta\to\infty \\ \im \zeta\ge\epsilon}}\zeta
  \sideset{}{'}\prod_{k\in M}
  \left(1-\frac{\zeta}{\mu_k}\right)
  \left(1-\frac{\zeta}{\lambda_k}\right)^{-1}\,,
  \quad\epsilon>0\,,
\end{equation*}
instead of (\ref{eq:limit-over-imaginary}).

\section*{Appendix: Boundary conditions for Jacobi
  operators}\sss \setcounter{section}{1}
\renewcommand{\theequation}{\Alph{section}.\arabic{equation}}
The difference expression $\gamma$ defined by
(\ref{eq:recurrence-spectral}) and (\ref{eq:initial-spectral})
can be written together in one equation with the help of some
conditions. Indeed, consider the difference expression
$\tilde\gamma$ given by
\begin{equation}
  \label{eq:general-difference}
   (\tilde\gamma f)_k = b_{k-1}f_{k-1} + q_k f_k + b_k
f_{k+1}\,,\quad k \in \mathbb{N}\quad (b_0=1)\,.
\end{equation}
Clearly, $\gamma f$ is equal to $\tilde\gamma f$ provided that
\begin{equation}
  \label{eq:Dir-bound-cond}
  f_0=0\,.
\end{equation}
This requirement can be considered as a boundary
condition for the difference equation
(\ref{eq:general-difference}). Notice that, although $f_0$ is
not an element of the sequence $\{f_k\}_{k=1}^\infty$, it can
be used to introduce boundary conditions for
(\ref{eq:general-difference}) which turn out to be completely
analogous to the boundary conditions at the origin for the
Sturm-Liouville operator on the semi-axis. We shall refer to
(\ref{eq:Dir-bound-cond}) as the Dirichlet boundary condition.
Thus, $J$ is the closure of the operator which acts on
sequences of $l_{fin}(\mathbb{N})$ by
(\ref{eq:general-difference}) with the Dirichlet boundary
condition (\ref{eq:Dir-bound-cond}).

Suppose that the deficiency indices of $J$ are $(1,1)$ and
consider now the following solution of
(\ref{eq:recurrence-spectral})
\begin{equation*}
   \tilde{v}_k(\beta):=Q_{k-1}(0)\cos\beta +
   P_{k-1}(0)\sin\beta\,,\quad\beta\in[0,\pi)\,.
\end{equation*}
Let us define the set
\begin{equation}
\label{eq:set-extensions}
  \bigl\{f=\{f_k\}_{k=1}^\infty\in l_2(\mathbb{N}):
  \tilde\gamma f\in l_2(\mathbb{N}),\,
  \lim_{n\to\infty}W_n(\tilde{v}(g),f)=0\bigr\}\,.
\end{equation}
Notice that $D(g)$, defined by
(\ref{eq:beta-extensions-domain}), coincides with
(\ref{eq:set-extensions}) as long as $g=\cot \beta$. As
pointed out in Section 2, the domain of every self-adjoint
extension of $J$ is given by (\ref{eq:set-extensions}) for
some $\beta$, and different $\beta$'s define different
self-adjoint extensions \cite{MR1711536}. Let us denote these
self-adjoint extensions by $J(g)$, as we did in Section 2,
bearing in mind that $g=\cot \beta$. The condition
\begin{equation}
  \label{eq:boundary-infinity}
  \lim_{n\to\infty}
  W_n(\tilde{v}(g),f)=0\,,
  \quad\ f\in\dom (J^*)
\end{equation}
determining the restriction of $J^*$ is considered to be a
boundary condition at infinity.

In analogy with the case of Sturm-Liouville operators one can
define general boundary conditions at zero for the difference
expression (\ref{eq:general-difference}). To this end, consider
the operator $J(\alpha,g)$ defined by the difference
expression (\ref{eq:general-difference}) with boundary
condition at infinity (\ref{eq:boundary-infinity}) if
necessary, and boundary condition ``at the origin''
\begin{equation}
    \label{eq:boundary-origin}
  f_1\cos\alpha +  f_0\sin\alpha =
  0\,,\qquad \alpha\in [0,\pi)\,.
\end{equation}
Thus, if $\alpha\in (0,\pi)$,
\begin{equation*}
    J(\alpha,g)=J(g)-\cot\alpha\langle\cdot,e_1\rangle e_1\,.
\end{equation*}
Therefore $J(\alpha,g)=J_h(g)$, provided that $h=\cot\alpha$.

When $\alpha=0$, from (\ref{eq:boundary-origin}), one has
$f_1=0$ and (\ref{eq:general-difference}) is used to define
the action of the operator for $k\ge 2$. $J(0,g)$ is said to
be operator $J(g)$ with Neumann boundary condition. For this
case we have that $J(0,g)$ is equal to $J_\infty(g)$.
\\[5mm]
\textbf{\large Acknowledgments}\hspace{3mm} We thank Rafael
del Rio for a hint on the literature.

\end{document}